\documentclass[12pt, hyper]{JHEP3}  
\pdfoutput=1
\usepackage{amsmath,amssymb,amsfonts,amsxtra, mathrsfs, makeidx,graphics,graphicx,amsthm,epsfig}

\newcommand{\todo}[1]{{\bf ?????!!!! #1 ?????!!!!}\marginpar{$\Longleftarrow$}}

\newcommand{\be}{\begin{equation}}
\newcommand{\ee}{\end{equation}}
\newcommand{\beq}{\begin{equation}}
\newcommand{\eeq}{\end{equation}}
\newcommand{\ba}{\begin{array}}
\newcommand{\ea}{\end{array}}
\newcommand{\bi}{\begin{itemize}}
\newcommand{\ei}{\end{itemize}}
\newcommand{\ben}{\begin{enumerate}}
\newcommand{\een}{\end{enumerate}}
\newcommand{\bea}{\begin{eqnarray}}
\newcommand{\eea}{\end{eqnarray}}
\newcommand{\bean}{\begin{eqnarray*}}
\newcommand{\eean}{\end{eqnarray*}}
\newcommand{\eref}[1]{(\ref{#1})}
\newcommand{\sref}[1]{\S\ref{#1}}
\newcommand{\tref}[1]{Table~\ref{#1}}
\newcommand{\fref}[1]{Figure~\ref{#1}}
\newcommand{\nn}{\nonumber}

\newcommand{\tr}{\mathop{\rm Tr}}

\newcommand{\BC}{\mathbb{C}}

\newcommand{\BZ}{\mathbb{Z}}

\newcommand{\cB}{{\cal B}}

\newcommand{\cF}{{\cal F}}

\newcommand{\cN}{{\cal N}}

\newcommand{\cG}{{\cal G}}
\newcommand{\cQ}{{\cal Q}}

\def\tcQ{\widetilde{\cQ}}

\newcommand{\PE}{\mathrm{PE}}
\newcommand{\PL}{\mathrm{PL}}

\newcommand{\ud}{\mathrm{d}}

\newcommand{\comment}[1]{}

\newcommand{\f}{{\cal F}^{\flat}}

\newcommand{\ggl}{{g}_{\mathrm{glue}}}

\newcommand{\eg}{\emph{e.g.}}
\newcommand{\ie}{\emph{i.e.}}

 \title{Tri-vertices and $SU(2)$'s}
\author{Amihay Hanany and Noppadol Mekareeya\\
Theoretical Physics Group, The Blackett Laboratory \\
Imperial College London, Prince Consort Road\\ 
London,  SW7 2AZ,  UK \\
Email: {\tt a.hanany, n.mekareeya07@imperial.ac.uk}}
\abstract{We examine a class of $\cN=2$ supersymmetric gauge theories in $(3+1)$ dimensions whose Lagrangians are determined by graphs consisting of two building blocks, namely a tri-vertex and a line. A line represents an $SU(2)$ gauge group and a tri-vertex represents a matter field in the trifundamental representation of $SU(2)^3$.  These graphs can be topologically classified by the genus and the number of external legs. This paper focuses on the hypermultiplet moduli spaces of the aforementioned theories.  We compute the Hilbert series which count all chiral operators on the hypermultiplet moduli space.  
Several examples show that theories corresponding to different graphs with the same genus and the same number of external legs possess the same Hilbert series.  This is in agreement with the conjecture that such theories are related to each other by S-duality.
We also give a general expression for the Hilbert series for the graph with any genus and any number of external legs.}
\preprint{Imperial/TP/10/AH/07}

\begin{document}
\section{Introduction}
Recently, a new class of $\cN=2$ superconformal field theories in $(3+1)$ dimensions has been explored \cite{Gaiotto:2009we}.  These theories are proposed to be the worldvolume theories of M5 branes wrapping Riemann surfaces.  In this paper, we focus on the case in which the number of M5 branes is two, so that the gauge groups involved are $SU(2)$'s.  These theories can be represented by graphs, called {\bf skeleton diagram}, consisting of lines and trivalent vertices, where a line represents an $SU(2)$ gauge group and a trivalent vertex represents a matter field in the tri-fundamental representation of $SU(2)^3$ (see \sref{sec:skel} for more details).  Such a graph defines a unique $\cN=2$ Langrangian in $(3+1)$ dimensions.  These graphs can be topologically classified by the genus $g$ and the number of external legs $e$.  

In this paper, we focus on the branch of the moduli space parametrised by the vacuum expectation values (VEVs) of the hypermultiplets (called the hypermultiplet moduli space or the {\bf Kibble branch}).  Certain quantities of the Kibble branch, such as dimension and some operators, of these theories or related ones have been discussed in, for example, \cite{Benini:2009gi, Benini:2009mz, Gadde:2010te, Nanopoulos:2010ga}.  In this paper, we compute the Hilbert series for the Kibble branch of various skeleton diagrams and show that {\it it is possible to count all chiral operators for any genus $g$ and any number of external legs $e$ of the skeleton diagram}.  This key result is explicitly stated in \eref{genge}.

The Hilbert series is a partition function for the chiral operators in the chiral ring of supersymmetric gauge theories.\footnote{There are also other similar quantities such as the superconformal index \cite{Kinney:2005ej, Romelsberger:2005eg, Dolan:2008qi, Gadde:2010te}, which is specific to superconformal field theories.  It would be interesting to find the relation between the Hilbert series and these quantities.} It can also be used as a primary tool to test various dualities in gauge theories, for example, in \cite{Benvenuti:2010pq} the Hilbert series is used in the context of the Argyres-Seiberg duality \cite{Argyres:2007cn}.  In this paper, several examples demonstrate that theories corresponding to different graphs with the same $g$ and $e$ possess the same Hilbert series.  This is in agreement with the conjecture that such theories are related to each other by S-duality \cite{Gaiotto:2009we}.

The outline and key results of this paper are as follows.  In \sref{sec:skel}, we summarise details of the skeleton diagram and give various simple examples.  In \sref{sec:kibble}, we introduce the notion of the Kibble branch of the moduli space and compute the dimension.  It is found that the dimension of the Kibble branch only depends on the external legs $e$ and not the genus $g$.  In \S\S \ref{sec:genuszero}, \ref{sec:genusone}, \ref{sec:zeroleg}, we compute Hilbert series for various examples.  The main results of this paper are collected in \sref{sec:main}.  These include the general formulae \eref{genge}, \eref{genfunc} and \eref{prodformgen}, which are a summary of all the results in this paper. 

\section{Skeleton diagrams of $\cN=2$ gauge theories} \label{sec:skel}
To write down a Lagrangian for a gauge theory with $\cN = 2$ supersymmetry it is sufficient to specify the gauge group, under which vector multiplets transform in the adjoint representation, and the representations under which the hypermultiplets transform.   In the case that hypermultiplets carry no more than two charges, it is convenient to represent the theory by a {\bf quiver diagram}, whose nodes and lines represent respectively vector multiplets and hypermultiplets.  Readers who are not familiar with $\cN=2$ quiver diagrams may wish to consult \cite{Benvenuti:2010pq} for further details.  However, when hypermultiplets carrying more than two charges, quiver diagrams are not good representatives of such theories.  Nevertheless, some of these theories can be represented graphically by {\bf skeleton diagrams}\footnote{These diagrams are also referred to as the `generalised quiver diagrams', first introduced in \cite{Gaiotto:2009we}.  In order to avoid a potential confusion with the notion of a quiver, we call such diagrams {\bf skeleton diagrams}.}, whose lines are assigned to the vector multiplets and vertices (or nodes) are assigned to hypermultiplets.   

This paper deals with an infinite class of $\cN=2$ supersymmetric gauge theories that are constructed by skeleton diagrams with the following simple rules:  The graphs are made out of lines and trivalent vertices. Each line (\includegraphics[height=0.05cm]{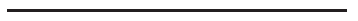}) represents an $SU(2)$ gauge group, with its length $L$ inversely proportional to its gauge coupling $g^2$, \ie~$L \sim 1/g^2$. (Therefore, a line with infinite length has zero coupling and therefore corresponds to a global $SU(2)$ symmetry.) Each tri-valent vertex (\includegraphics[height=0.5cm]{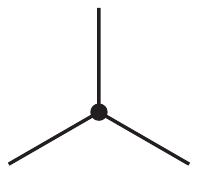}) represents a half-hypermultiplet $\cQ_{\alpha \beta \gamma}$ transforming in the  $[1;1;1]$ representation\footnote{In this paper, we denote irreducible representations and their characters by the Dynkin labels (which are highest weights of the corresponding representations). For example, for $SU(2)$, $[1]$ denotes the two-dimensional (fundamental) representation, and $[2]$ denotes the three-dimensional (adjoint) representation. In the case of product groups, we use $;$ to separate the highest weights of the representations from different groups. For example, $[1;1;1]$ of $SU(2)^3$ denotes the tri-fundamental representation of $SU(2)^3$.} of $SU(2)^3$, where the indices $\alpha,~\beta,~\gamma = 1, 2$ corresponds to three different $SU(2)$ groups.  A skeleton diagram defines a unique $\cN=2$ Langrangian in $(3+1)$ dimensions.

In the $\cN=1$ language, each $\cN=2$ vector multiplet decomposes into an $\cN=1$ vector multiplet and an $\cN=1$ chiral multiplet.  Each $\cN=2$ half-hypermultiplet decomposes into an $\cN=1$ chiral multiplet. Finally, the superpotential takes the form of a sum over all nodes with a contribution of each node is
\bea
\cQ_{\alpha \beta \gamma} \cQ_{\alpha' \beta' \gamma'} \left(\phi_1^{\alpha \alpha'} \epsilon^{\beta \beta'} \epsilon^{\gamma \gamma'}+ \epsilon^{\alpha \alpha'} \phi_2^{\beta \beta'} \epsilon^{\gamma \gamma'} + \epsilon^{\alpha \alpha'} \epsilon^{\beta \beta'} \phi_3^{\gamma \gamma'} \right)~, \label{gensup}
\eea
where the three sets of indices $\{ \alpha, \alpha' =1,2\}, \{ \beta, \beta' =1,2\}, \{\gamma, \gamma' =1,2\}$ correspond to the three different $SU(2)$ groups, and the adjoint chiral multiplets $\phi_1, \phi_2, \phi_3$ come from the three different $SU(2)$ $
\cN=2$ vector multiplets.  By convention, infinite lines give rise to adjoint valued mass terms.  Note that the superpotential \eref{gensup} is defined up to a constant which is determined by the $\cN=2$ supersymmetry

A motivation of skeleton diagrams comes from the study of $\cN=2$ supersymmetric gauge theories living on M5-branes wrapping Riemann surfaces \cite{Gaiotto:2009we, Gaiotto:2009gz}.  In this paper, we focus on the theories with $SU(2)$ symmetries, and so the number of M5-branes involved is two.  The topology of the skeleton diagram is the same as that of the corresponding Riemann surface, namely the number of loops of the skeleton diagram is the genus of the Riemann surface and the number of external legs of the skeleton diagram is the number of punctures on the Riemann surface.  

Below we give a few examples of the $\cN=2$ theories with their skeleton diagrams.

\subsection{The theory with a free trifundamental of $SU(2)^3$} 
\begin{figure}[htbp]
\begin{center}
\includegraphics[height=2.5cm]{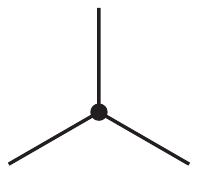}
\caption{The theory with a free trifundamental field of $SU(2)^3$.}
\label{trifund}
\end{center}
\end{figure}

Let us consider the theory with a tri-vertex and three external legs (\fref{trifund}).  Each of the three legs corresponds to an $SU(2)$ global symmetry.  The vertex corresponds to 8 free, possibly massive, half-hypermultiplets $\cQ_{ijk}$ transforming in the trifundamental $[1;1;1]$ representation of the $SU(2)^3$ global symmetry.  The possible mass terms are
\bea
W = \cQ_{ijk} \cQ_{i' j' k'} \left(m_1^{i i'} \epsilon^{j j'} \epsilon^{k k'}+ \epsilon^{i i'} m_2^{j j'} \epsilon^{k k'} + \epsilon^{i i'} \epsilon^{j j'} m_3^{k k'} \right)~,
\eea
where $i,~j,~k,~i',~j',~k' = 1,2$.
This theory is also known in the literature as {\bf the $T_2$ theory}.  Subsequently, we use this theory as a building block to construct a number of other theories by means of `gluing'.

\subsection{The $SU(2)$ $\cN=4$ gauge theory with two free singlets} 
\begin{figure}[htbp]
\begin{center}
\includegraphics[height=2cm]{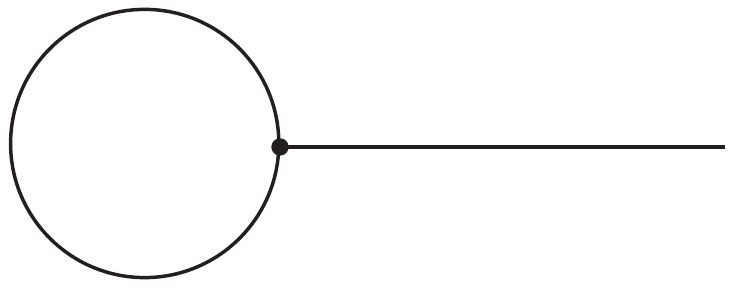}
\caption{(The tadpole) The $SU(2)$ $\cN=4$ gauge theory with two singlets.}
\label{N4SYM}
\end{center}
\end{figure}

Let us consider the tadpole diagram in \fref{N4SYM}.  This diagram can be obtained by gluing together two external legs in a $T_2$ theory. As shown in the diagram, this theory has an $SU(2)$ gauge group (corresponding to the loop) and an $SU(2)$ global symmetry (corresponding to the external leg).  

The vertex corresponds to a half-hypermultiplet $\cQ_{abi}$, where $a,b=1,2$ are $SU(2)$ gauge indices and $i =1,2$ is an $SU(2)$ global index. 
Let us define the trace of $\cQ$ and the traceless part of $\cQ$ as
\bea
X_{i} &\equiv& \epsilon^{ab} \cQ_{abi}~, \nn \\
\varphi_{abi} &\equiv& \cQ_{abi} -\frac{1}{2} X_{i} \epsilon_{ab}~.
\eea
Note that, by definition, the half-hypermultiplets $\varphi_i$ are traceless, \ie~$\epsilon^{ab} \varphi_{abi} = 0$. Hence, $\varphi$ is an $SU(2)$ adjoint hypermultiplet.  The vector multiplet of the $SU(2)$ gauge group and the adjoint hypermultiplet $\varphi$ give rise to an $\cN=4$ gauge theory with an $SU(2)$ gauge group.  

On the other hand, the gauge singlet $X$ is a free hypermultiplet which is more conveniently written as two half-hypermultiplets $X_1, X_2$ transforming in the fundamental representation of the $SU(2)$ global symmetry.  

There is also a global $U(1)$ $R$-symmetry under which the half-hypermultiplets $
\cQ_{ab, \alpha}$ carries the charge $1$ (which is also the scaling dimension).   

The representations in which $X$ and $\varphi$ transform are summarised in \tref{matterN4sym}.


\begin{table}[htdp]
\begin{center}
\begin{tabular}{|c||c||c|c|}
\hline
Field & Gauge $SU(2)$ & Global $SU(2)$ & Global $U(1)$ \\
\hline
Fugacity: & $z$ & $x$ & $t$ \\
\hline
$\varphi$ & $[2]$ & $[1]$ & 1 \\
$X$  & $[0]$ & $[1]$ &1 \\
\hline
\end{tabular}
\caption{The hypermultiplets in the tadpole theory.}
\label{matterN4sym}
\end{center}
\end{table}%

\comment{
}

Let $\phi$ be the scalar field in the $\cN=2$ $SU(2)$ vector multiplet. In an $\cN=1$ supersymmetric notation, one can write down the superpotential \eref{gensup}, including a mass term, as 
\bea
W &=& \cQ_{abi} \cQ_{a' b'j} \left( \phi^{a a'} \epsilon^{b b'} \epsilon^{ij} + \epsilon^{a a'} \phi^{b b'} \epsilon^{ij} + \epsilon^{a a'} \epsilon^{b b'} m^{ij} \right)~. \nn
\eea
For simplicity, we set the mass term to zero and obtain
\bea
W =  \cQ_{ab i} \cQ_{a' b'j} \left( \phi^{a a'} \epsilon^{b b'} \epsilon^{ij} + \epsilon^{a a'} \phi^{b b'} \epsilon^{ij} \right)~.
\eea  
Observe that, by symmetry, the trace $\epsilon^{ab} \cQ_{abi} = X_i$ does not contribute to the superpotential. Indeed, $X$ is a free hypermultiplet.  One can therefore write down the above superpotential using the traceless part of $\cQ$ as
\bea
W = 2\phi^{a a'} \epsilon^{b b'} \epsilon^{ij} \varphi_{abi} \varphi_{a' b'j}~.  \nn
\eea
Note that the factor in front of the superpotential is determined by supersymmetry but is not relevant to the computations done in this paper.  We shall henceforth drop this factor and take
\bea
W = \phi^{a a'} \epsilon^{b b'} \epsilon^{ij} \varphi_{abi} \varphi_{a' b'j}~. \label{suptadphi}
\eea

\subsection{$SU(2)$ gauge theory with 4 flavours} 
\begin{figure}[htdp]
\begin{center}
\includegraphics[height=2.5cm]{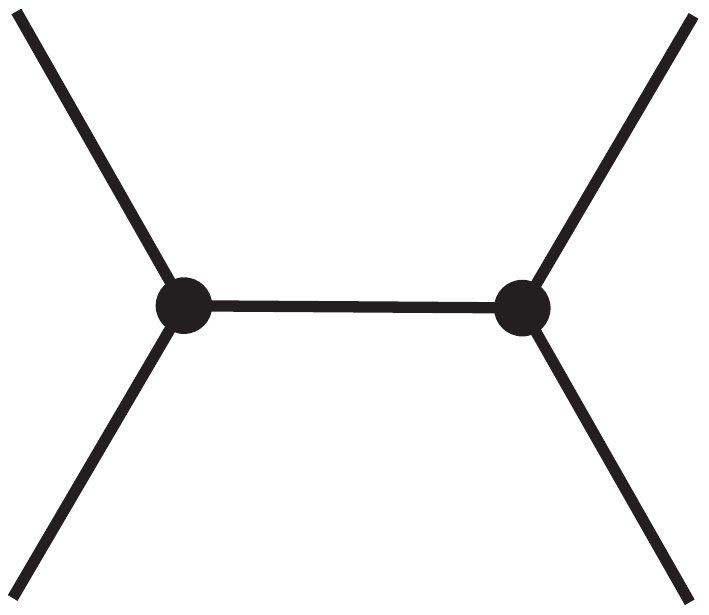}
\caption{The skeleton diagram of the $SU(2)$ gauge theory with 4 flavours.}
\label{su2w4flv}
\end{center}
\end{figure}

Consider the skeleton diagram in \fref{su2w4flv}.  This diagram can be obtained by `gluing' two $T_2$ theories along one of the external legs in each diagram.

The internal line corresponds to the $SU(2)$ gauge group. Each of the four external legs corresponds to an $SU(2)$ global symmetry.  
The two nodes represent two trifundamental fields $\cQ_{i_1 i_2  a}$ and $\tcQ_{i_3 i_4 a}$ of $SU(2)^3$, where $a$ is an $SU(2)$ gauge index and $i_1, i_2, i_3, i_4$ are the indices for the four different $SU(2)$ flavour symmetries.

Let $\phi$ be a scalar field in the $\cN=2$ $SU(2)$ vector multiplet.
In an $\cN=1$ supersymmetric language, the superpotential (with mass terms) can be written as
\bea \label{su24flv1}
W &=& \cQ_{i_1 i_2 a} \cQ_{i_1' i_2' a'} \left(m_1^{i_1 i_1'} \epsilon^{i_2 i_2'} \epsilon^{a a'}+ \epsilon^{i_1 i_1'} m_2^{i_2 i_2'} \epsilon^{a a'} + \epsilon^{i_1 i_1'} \epsilon^{i_2 i_2'} \phi^{a a'} \right) \nn \\
&& + \tcQ_{i_3 i_4 a} \tcQ_{i_3' i_4' a'} \left(m_3^{i_3 i_3'} \epsilon^{i_4 i_4'} \epsilon^{a a'}+ \epsilon^{i_3 i_3'} m_4^{i_4 i_4'} \epsilon^{a a'} + \epsilon^{i_3 i_3'} \epsilon^{i_4 i_4'} \phi^{a a'} \right)~.
\eea

\begin{figure}[htdp]
\begin{center}
\includegraphics[height=1.5cm]{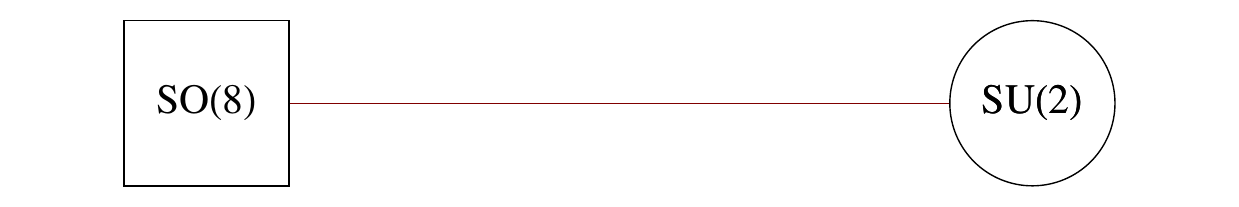}
\caption{The quiver diagram of the $SU(2)$ gauge theory with 4 flavours.}
\label{pic:su2w4flvquiv}
\end{center}
\end{figure}

From the following decompositions of $SO(8)$ into $SU(2)^4$:
\bea
\left[1,0,0,0\right]_{SO(8)} = [1;1;0;0]+[0;0;1;1]~,
\eea
one can combine $(i_1, i_2), (i_3, i_4)$ into one $SO(8)$ index $I$, and hence the $SU(2)^4$ global symmetry enhances to $SO(8)$.  The 16 half-hypermultiplets $\cQ_{i_1 i_2 a}$ and $\tcQ_{i_3 i_4 a}$ can then be combined into $Q_{aI}$, which are indeed the quarks in an $SU(2)$ gauge theory with 4 flavours.  The quiver diagram of this theory is depicted in \fref{pic:su2w4flvquiv}.

In $\cN=1$ supersymmetric notation, one can rewrite the superpotential as
\bea \label{su24flv2}
W= Q_{aI} Q_{bI} \phi^{ab} + \mu^{IJ} Q_{aI} Q_{bJ} \epsilon^{ab}~.
\eea
Let us compare \eref{su24flv1} with \eref{su24flv2}.  The mass parameters $\mu_{IJ}$ transform in the adjoint representation $[0,1,0,0]$ of $SO(8)$.  This can be decomposed into $SU(2)^4$ representations as
\bea
[0,1,0,0]_{SO(8)} = [1;1;1;1]+[2;0;0;0]+[0;2;0;0]+[0;0;2;0]+[0;0;0;2]~.
\eea
We see from \eref{su24flv1} that the mass parameters $m_1^{i_1 i_1'}$, $m_2^{i_2 i_2'}$, $m_3^{i_3 i_3'}$ and $m_4^{i_4 i_4'}$ transform respectively in the $SU(2)^4$ representations $[2;0;0;0]$, $[0;2;0;0]$, $[0;0;2;0]$ and $[0;0;0;2]$.  Therefore, we have the following tensor decomposition:
\bea
\mu^{IJ}~\rightarrow ~ m^{i_1 i_2 i_3 i_4} + m_1^{i_1 i_1'}+ m_2^{i_2 i_2'}+ m_3^{i_3 i_3'}+ m_4^{i_4 i_4'}~,
\eea
where the mass parameters $m^{i_1 i_2 i_3 i_4}$ transform in $[1;1;1;1]$ of $SU(2)^4$.
Observe that we can set $m^{i_1 i_2 i_3 i_4}$ to zero by an $SO(8)$ transformation.

\section{The Kibble branch of the moduli space} \label{sec:kibble}
\paragraph{Topology of the skeleton diagram.} One can classify the skeleton diagrams according to their topological properties, namely the genus $g$ and the number of external legs $e$.  Henceforth, we collect these numbers in an ordered pair $(g,e)$.  Given $g$ and $e$, the number of internal lines is $3g-3+e$ and the number of nodes (and also the number of $T_2$ building blocks) is the Euler characteristic $\chi = 2g-2+e$.  Recall that an internal line corresponds to a gauge group and each node corresponds to a trifundamental matter field.  Therefore,
\bea
\text{The number of $SU(2)$ gauge groups} = \cG(g,e) &=& 3g-3+e~, \nn \\
\text{The number of matter fields} = \chi(g,e) &=& 2g-2+e~, \nn\\
\text{The number of $SU(2)$ global symmetries} &=& e~. \label{Gchi}
\eea

In $\cN=2$ supersymmetric gauge theories with one gauge group, one typically refers to two branches of the moduli space, namely the Higgs branch and the Coulomb branch.  The Higgs branch is the branch on which the gauge group is {\it completely} broken and the vector multiplet becomes massive via the Higgs mechanism; this branch is parametrised by the massless gauge singlets of hypermultiplets.  The Coulomb branch is, on the other hand, the branch on which the gauge group is broken to a collection of $U(1)$'s and the hypermultiplets generically become massive; this branch is parametrised by complex scalars in the vector multiplet.  

However, for the theories with genus $g \geq 1$, the gauge group is not completely broken on the branch which is parametrised by VEVs of hypermultiplets.   We conjecture that at a generic point in this branch the $SU(2)^\cG$ gauge symmetry is broken to $U(1)^g$ (see Appendix \ref{sec:unbrokenu1}).  In order to avoid a potential confusion with the notion of Higgs branch, we refer to this branch of the moduli space as the {\bf Kibble branch}\footnote{In honour of Professor Tom Kibble's contribution to the theory of spontaneous symmetry breaking.}, denoted by $\mathcal{K}$.  Note however that for theories with zero genus $g=0$, the Kibble branch coincides with the Higgs branch.

Let us compute the dimension of the Kibble branch for theories with genus $g$ and $e$ external legs. Since each $T_2$ building block contains $8$ half-hypermultiplets (or equivalently $4$ hypermultiplets) and there are $\chi$ of such building blocks, the hypermultiplets have $4 \chi$ quaternionic degrees of freedom in total.  At a generic point on the Kibble branch the $SU(2)^\cG$ gauge symmetry is broken to $U(1)^g$, and hence there are $3 \cG-g$ broken generators.  As a result of the Higgs mechanism, the vector multiplet gains $3 \cG-g$ quarternionic degrees of freedom and become a massive $\cN=2$ vector multiplet.  Thus, from \eref{Gchi}, the $4\chi - (3 \cG-g) = e+1$ quarternionic degrees of freedom are left massless.  Thus, the quarternionic dimension of the Kibble branch is
\bea
\dim_{\mathbb{H}} \mathcal{K} = e+1~. \label{dimmod}
\eea
This is in agreement with \cite{Benini:2009gi}.
Note that the dimension of the Kibble branch does not depend on the genus, but depends only on the number of external legs. 

\section{Theories with genus zero} \label{sec:genuszero}
In this section, we focus on the Hilbert series of theories with genus zero.  Below the Hilbert series of these theories are studied in detail.

\subsection{The $T_2$ theory $(g=0, e=3)$}
It is clear that the moduli space of the $T_2$ theory is generated by the trifundamental field.  Hence, the operators transform in the symmetric powers of $[1;1;1]$ of $SU(2)^3$.  Thus, the Hilbert series of this theory can be written in an elegant way using the plethystic exponential ($\PE$)
\bea
g_{T_2} (t; x_1,x_2,x_3) =\PE \left[ [1;1;1] t \right] =  \prod_{
\epsilon_i=
\pm1} \frac{1}{1-t x_1^{\epsilon_1} x_2^{\epsilon_2} x_3^{\epsilon_3}}~, \label{T2PE}
\eea
where $x_1$, $x_2$ and $x_3$ are the fugacities of $SU(2) \times SU(2) \times SU(2)$, and the plethystic exponential $\PE$ of a multi-variable function $f(t_1, . . . , t_n)$ that vanishes at the origin, $f(0,...,0) = 0$, is defined as
\bea
\PE \left[ f(t_1, t_2, \ldots, t_n) \right] = \exp \left( \sum_{k=1}^\infty \frac{1}{k} f(t_1^k, \ldots, t_n^k) \right)~.
\eea
This expression \eref{T2PE} is manifestly symmetric under any permutation of the 3 external legs. The permutation group $S_3$ acts on exchanging the legs and the Hilbert series on the Kibble branch is an invariant function of this $S_3$.
This point is used below to demonstrate the invariance of the Hilbert series on the Kibble branch.

One can rewrite \eref{T2PE} in terms of infinite sums of the irreducible representations of $SU(2)^3$ as
\bea
 g_{T_2} (t; x_1,x_2,x_3) &=& \frac{1}{1-t^4} \sum_{n_1, n_2 ,n_3,m = 0}^\infty  \left( [2n_1+m; 2n_2+m; 2n_3+m] t^{2n_1+2n_2+2n_3+m} + \right. \nn \\
&& \left. [2n_1+m+1; 2n_2+m+1; 2n_3+m+1] t^{2n_1+2n_2+2n_3+m+3}  \right)~. \label{T2}
\eea
As is shown below, this infinite sum turns out to be more useful for generalisation to any pair $(g,e)$.  In this expression, there are 4 sums, one for each external leg, and one that `glues' all expressions together (without it, the sums would simply factorise).

\subsection{$SU(2)$ gauge theory with $4$ flavours $(g=0, e=4)$}
The Hilbert series of this theory is computed in (4.12) of \cite{Benvenuti:2010pq}.  In terms of $SO(8)$ representations, this can be written as
\bea
g_{N_c=2, N_f=4} (t, z_1, z_2, z_3, z_4) = \sum_{k=0}^\infty [0,k,0,0]_{SO(8)} t^{2k}~, \label{so8}
\eea 
where $z_1, z_2, z_3, z_4$ are the $SO(8)$ fugacities.
The moduli space of this theory is 10 complex dimensional (see \eg, \S4.2 of \cite{Benvenuti:2010pq}).  This is in agreement with \eref{dimmod}.  

\paragraph{A branching rule of $SO(8)$ to $SU(2)^4$.}  Let us decompose these $SO(8)$ representations into $SU(2)^4$ representations.  A map from the $SO(8)$ fugacities to the $SU(2)^4$ can be chosen to be
\bea
z_1 = x_1 x_2,\quad z_2 = x_2^2,\quad z_3 = x_3 x_2,\quad z_4 = x_4 x_2~,
\eea
where $x_1, x_2, x_3, x_4$ are the four $SU(2)$ fugacities.  With such a map, one obtains, \eg
\bea
[1,0,0,0]_{SO(8)} &=& [1;1;0;0] + [0;0;1;1]~, \nn \\
\left[0,1,0,0 \right]_{SO(8)} &=& [1;1;1;1] + [2;0;0;0] + [0;2;0;0] + [0;0;2;0] + [0;0;0;2]~, \quad
\eea
etc.
The formula \eref{so8} can be rewritten in terms of $SU(2)$ representations as
\bea
&& g_{N_c=2, N_f=4} = \frac{1}{1-t^4} \sum_{n_1, \ldots ,n_4,m = 0}^\infty  \left( [2n_1+m; 2n_2+m; 2n_3+m; 2n_4+m] t^{2n_1+2n_2+2n_3+2n_4+2 m} + \right. \nn \\
&& \left. [2n_1+m+1; 2n_2+m+1; 2n_3+m+1; 2n_4+m+1] t^{2n_1+2n_2+2n_3+2n_4+2m+4}  \right)~. \label{su24flv}
\eea
This is a form, which as in \eref{T2}, turns out to be the right form to generalise to any pair $(g,e)$.  This expression is invariant under any permutation of the external legs. The permutation group $S_4$ acts on exchanging the legs and the Hilbert series on the Kibble branch is an invariant function of this $S_4$.

\subsubsection{Gluing two $T_2$ theories.}
One can also obtain the $SU(2)$ gauge theory with 4 flavours by gluing two $T_2$ theories along the external legs.  Before obtaining the Hilbert series, let us briefly summarise the gluing technique.

\subsubsection*{A summary of the gluing technique} 
In \cite{Benvenuti:2010pq}, we derive Hilbert series when two Riemann surfaces are glued together along the punctures.  Let us briefly summarise the gluing procedure.  Suppose that the maximal punctures along which we glue possess the symmetry of a group $G$, whose fugacites are denoted collectively by $z_k$. Let the Hilbert series of the theory corresponding to the first Riemann surface be $g_1 (t, x_i, z_k)$ and let the one corresponding to the second Riemann surface be $g_2(t, y_j, z_k)$, where $x_i, y_j$ represent a dependence on additional fugacities.  The Hilbert series when two Riemann surfaces are glued together is given by
\bea
g(t,x_i, y_j) = \int \ud \mu_G (z_k)~g_1 (t, x_i, z_k) ~ g_{\mathrm{glue}}(t,z_k) ~ g_2 (t, y_j, z_k)~, \label{gluefor}
\eea
where the gluing factor (when there is no `self-gluing' involved) is 
\bea
g_{\mathrm{glue}}(t,z_k) = \frac{1}{\PE \left[ Adj(z_k) t^2 \right]}~. \label{gluingfac}
\eea
In particular, for the $SU(2)$ group, the gluing factor is given by
\bea
g_{\mathrm{glue}}(t,z) = \frac{1}{\PE \left[ [2]_z t^2 \right]} = 1 -t^2 [2]_z + t^4 [2]_z -t^6~, \label{gfsu2}
\eea
where $[2]_z = z^2 +1 + \frac{1}{z^2} $. 

Note however that, when the gluing involves self-gluing of the Riemann surface, the gluing does not take the form \eref{gluingfac}.  We demonstrate this point in \sref{selfglue1}.

\comment{
\paragraph{Gluing technique and character expansions.} In this paper, our main interests involve the $SU(2)$ group.  In many cases, the Hilbert series $g_1$ and $g_2$ can be written as infinite sums over the irreducible representations of a product of $SU(2)$, such as in \eref{su24flv}. Let us suppose that the $SU(2)$ representations associated with the gluing external legs in $g_1$ is $[r_1]$ and in $g_2$ is $[r_2]$.  Recalling the orthogonality relation
\bea
\int \ud \mu_{SU(2)} [p] [q] = \delta_{p,q}~,
\eea
we can rewrite \eref{gluefor} as
\bea
g(t, a, b) = \sum_{a,b} \sum_{r_1=0}^\infty \sum_{r_2 = 0}^\infty g_{\mathrm{glue}}(r_1, r_2) g_1 (t,a, r_1) g_2 (t,b, r_2)~, \label{charglue}
\eea
where $a$ and $b$ denote collectively the representations which do not involve $[r_1]$ and $[r_2]$, and the gluing factor \eref{gluingfac} can be written as
\bea
g_{\mathrm{glue}}(r_1, r_2) = \left \{
\begin{array}{ll}
1-t^6 & ~\text{if } r_1 = r_2 =0~, \\
(1-t^6) \delta_{r_1,r_2}- t^2(1-t^2) (\delta_{r_1,r_2} +\delta_{r_1,r_2+2}  + \delta_{r_1+2,r_2} ) & ~\text{otherwise}~. \\
\end{array} \right. \nn \\
\eea
}

\subsubsection*{The $SU(2)$ theory with 4 flavours - revisited} 
Let us suppose that the legs $3$ (associated with the fugacity $z$) of the two $T_2$ are glued together. To obtain the Hilbert series, we apply the gluing formula \eref{gluefor} to the Hilbert series \eref{T2} of $T_2$:
\bea
\int \ud \mu_{SU(2)}(z) ~g_{T_2} (t; x_1, x_2,z) ~g_{\mathrm{glue}}(t,z) ~g_{T_2} (t; x_3, x_4,z)~.
\eea
The gluing factor and the integration impose a `selection rule' on $m$.  In order to determine which $m$ survive, we use the following identities:
\bea
\int \ud \mu_{SU(2)}(z) \sum_{n_3, m, n'_3, m'} g_{\mathrm{glue}}(t,z)  [2n_3+m]_z [2n'_3+m']_z~  t^{2n_3+m+2n'_3+m'} &=& 1+t^2-t^4 \nn \\
\int \ud \mu_{SU(2)}(z) \sum_{n_3, m, n'_3, m'} g_{\mathrm{glue}}(t,z)  [2n_3+m]_z [2n'_3+m'+1]_z t^{2n_3+m+2n'_3+m'+3} &=& t^4  \nn \\
\int \ud \mu_{SU(2)}(z) \sum_{n_3, m, n'_3, m'} g_{\mathrm{glue}}(t,z) [2n_3+m+1]_z [2n'_3+m']_z t^{2n_3+m+2n'_3+m'+3} &=& t^4  \nn \\
\int \ud \mu_{SU(2)}(z) \sum_{n_3, m, n'_3, m'} g_{\mathrm{glue}}(t,z)  [2n_3+m+1]_z [2n'_3+m'+1]_z t^{2n_3+m+2n'_3+m'+6} &=& t^6~, \nn \\ \label{iden1}
\eea 
where the summations are over $n_3, m, n'_3, m'$ from $0$ to $\infty$ and the subscripts $z$ indicates that the characters depend on $z$.
The first and the second identities contribute to the first term in \eref{su24flv}:
\bea
\frac{1+t^2-t^4}{1-t^4} + \frac{t^4}{1-t^4} = \frac{1+t^2}{1-t^4} =\frac{1}{1-t^2} = \sum_{m=0}^\infty t^{2m}~. \label{contr1}
\eea
The third and the fourth identities contribute to the second term in \eref{su24flv}:
\bea
\frac{t^4}{1-t^4} + \frac{t^6}{1-t^4} = \frac{t^4}{1-t^2} = \sum_{m=0}^\infty t^{2m+4}~. \label{contr2}
\eea
Hence, we arrive at \eref{su24flv}, as expected.

\subsubsection*{Derivation of the identities \\ {\small (The reader may skip this topic without the loss of continuity.)}} 
We discuss the derivation of the first identity in \eref{iden1}; the others can be derived in a similar fashion.  Let us first focus on the following expression:
\bea
\mathcal{A} &\equiv& \int \ud \mu_{SU(2)}(z) \sum_{n_3, m, n'_3, m'}  [2n_3+m]_z [2n'_3+m']_z t^{2n_3+m+2n'_3+m'} \nn \\
&=&  \sum_{n_3, m, n'_3, m'} \delta_{2n_3+m, 2n'_3+m'} t^{2n_3+m+2n'_3+m'}~.
\eea
For a given value of $2n_3+m$, there are $\lfloor (2n_3+m)/2+1\rfloor = n_3+1+ \lfloor m/2 \rfloor$ pairs of $(n'_3, m')$ which give non-zero delta functions. Therefore, we have 
\bea
\mathcal{A}=\sum_{n_3, m, n'_3, m'} \delta_{2n_3+m, 2n'_3+m'} t^{2n_3+m+2n'_3+m'}  
&=&  \sum_{n_3=0}^\infty  \sum_{k=0}^\infty (n_3+1+ \lfloor m/2 \rfloor) t^{4n_3+2m}  \nn \\
&=& \sum_{n_3=0}^\infty  \sum_{k=0}^\infty (n_3+1+k) t^{4n_3+4k}(1+t^2) \nn \\
&=& \frac{1+t^4}{(1-t^4)^2(1-t^2)}~,
\eea
where, in the second line, we considered the two separated cases, $m = 2k$ and $m =2k+1$.
Next, we consider the expression
\bea
\mathcal{B} &\equiv& \int \ud \mu_{SU(2)}(z) \sum_{n_3, m, n'_3, m'} [2]_z [2n_3+m]_z [2n'_3+m']_z t^{2n_3+m+2n'_3+m'} \nn \\
&=&  -1+\sum_{n_3, m, n'_3, m'} (\delta_{2n_3+m, 2n'_3+m'}+\delta_{2n_3+m+2, 2n'_3+m'}+\delta_{2n_3+m, 2n'_3+m'+2}) t^{2n_3+m+2n'_3+m'} \nn\\
&=&  -1+\frac{1+t^4}{(1-t^4)^2(1-t^2)}+ 2 \times \frac{2 t^2}{\left(1-t^2\right)^3 \left(1+t^2\right)^2} \nn \\
&=& \frac{t^2 \left(5+3 t^2-2 t^4-t^6+t^8\right)}{\left(1-t^2\right)^3 \left(1+t^2\right)^2}~,
\eea
where $-1$ in the second line compensate the case in which $2n_3+m = 2n'_3+m' =0$.
Using the gluing factor \eref{gluingfac} (with $Adj=[2]$), we find that
\bea
&& \int \ud \mu_{SU(2)}(z) \sum_{n_3, m, n'_3, m'} g_{\mathrm{glue}}(t,z)  [2n_3+m]_z [2n'_3+m']_z~  t^{2n_3+m+2n'_3+m'} \nn \\
&=& (1-t^6) \mathcal{A} - t^2(1-t^2) \mathcal{B} = 1 + t^2 - t^4~.
\eea

\subsubsection{Three phases of $SU(2)$ theory with 4 flavours} 
\begin{figure}[htbp]
\begin{center}
\includegraphics[height=3 cm]{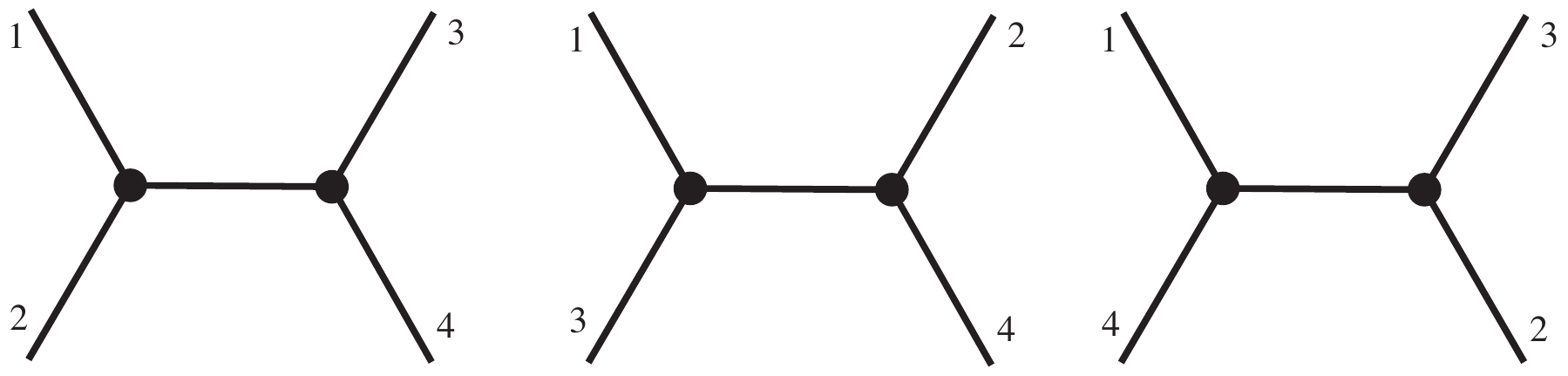}
\caption{The three weak coupling limits of an $SU(2)$ gauge theory with 4 flavours.}
\label{threephases}
\end{center}
\end{figure}

As pointed out in \cite{Gaiotto:2009we}, there are 3 weak coupling limits of an $SU(2)$ gauge theory with 4 flavours.  These corresponds to the permutations of the labels of the external legs (depicted in \fref{threephases}).  They have different origins from the perspective of theories on M5-branes wrapping Riemann surfaces.  For example, the theory at the centre of \fref{threephases} can be obtained from the gluing of two Riemann surfaces; one contains punctures $1$ and $3$ and the other contains puctures $2$ and $4$.  All of these phases are conjectured to be related to each other by S-duality \cite{Gaiotto:2009we} which states that the IR dynamics of these theories are identical.  Indeed, it can easily be seen from \eref{su24flv} that the Hilbert series of these three phases are identical, since the permutations of the labels correspond to the permutations of $n_1, \ldots, n_4$, the dummy variables in the summations.

In fact, a stronger version of this duality is that the IR dynamics depends on the pair $(g,e)$ only and not on the specific choice of the Lagrangian. A consistency check of this duality is the set of computations below which demonstrate that the Hilbert series on the Kibble branch is an invariant of S-duality, or alternatively, depends on the choice of the pair $(g,e)$ and not on other details of the skeleton diagram.


\section{Theories with genus one} \label{sec:genusone}
 Below the Hilbert series of theories with genus one are studied in detail.
 
\subsection{The tadpole theory $(g=1, e=1)$}
In this subsection, we compute the Hilbert series of the Kibble branch of the tadpole theory (\fref{N4SYM}).  We translate the $\cN=2$ data into the $\cN=1$ language.  Let us denote the scalar in the vector multiplet by $\phi$.  In the $\cN=1$ language, the superpotential can be written as
\bea
W = \cQ_{ab i} \cQ_{a' b' j} \left( \phi^{a a'} \epsilon^{b b'} \epsilon^{i j} + \epsilon^{a a'} \phi^{b b'} \epsilon^{i j} \right)~.
\eea 
On the Kibble branch, the field $\phi$ becomes massive and hence $\langle \phi \rangle = 0$.  Therefore, the non-trivial F-terms are
\bea
(\cQ_{ab i} \cQ_{a' b' j}   + \cQ_{b a i} \cQ_{b' a' j}) \epsilon^{b b'} \epsilon^{i j}= 0~.
\eea
Using the fugacities according to \tref{matterN4sym}, the Hilbert series of the two commuting adjoint fields is 
\bea
&& (1-t^2[2;0]+t^3[0;1]) \PE \left[ [2;1] t \right]  \nn \\
&=& \frac{1-t^2(1+z^2+\frac{1}{z^2})+t^3 (x+\frac{1}{x})}{\left(1-\frac{t}{x}\right) (1-t x)\left(1-\frac{t}{x z^2}\right) \left(1-\frac{t x}{z^2}\right) \left(1-\frac{t z^2}{x}\right) \left(1-t x z^2\right)}~,
\eea
where $[a; b]$ denotes the product of the characters $[a]_z [b]_x$.
The F-flat Hilbert series is then given by
\bea
\f(t, z, x) &=&  (1-t^2[2;0]+t^3[0;1]) \PE \left[ [2;1] t +[0;1] t \right] \nn \\
&=& \frac{1-t^2(1+z^2+\frac{1}{z^2})+t^3 (x+\frac{1}{x})}{\left(1-\frac{t}{x}\right)^2 (1-t x)^2 \left(1-\frac{t}{x z^2}\right) \left(1-\frac{t x}{z^2}\right) \left(1-\frac{t z^2}{x}\right) \left(1-t x z^2\right)}~. \label{fftad}
\eea
Integrating over the $SU(2)$ gauge group, one obtains the Kibble branch Hilbert series
\bea
g_{\mathrm{tadpole}}(t,x) &=& \frac{1-t^4}{(1-t x)(1- \frac{t}{x})(1-t^2)(1-t^2 x^2)(1-\frac{t^2}{x^2}) } \nn \\
&=& (1-t^4) \PE \left[ [1] t + [2] t^2 \right] \nn \\
&=& \frac{1}{1-t^4} \sum_{n_1, n_2,m = 0}^\infty \left( [2n_1+m] t^{2n_1+ m} + [2n_1+m+1] t^{2n_1+2n_2+m+3} \right) ~. \nn \\ \label{HStad}
\eea
The Kibble branch is therefore a 4 complex dimensional complete intersection.  The generators are $X$ at order $t$ and 
\bea
M_{i j} = \epsilon^{a a'} \epsilon^{b b'}  \cQ_{abi} \cQ_{a' b'j} \label{Mtadpole}
\eea
at order $t^2$.  The relation at order $t^4$ is
\bea
\det M = 0~. \label{reltadpole}
\eea

Note that the Kibble branch is actually $\BC^2/\BZ_2 \times \BC^2 $, where $\BC^2/\BZ_2$ is generated by $M_{\alpha \beta}$ and $\BC^2$ is generated by the two gauge singlet $X_\alpha$.  This can also be seen from the fact that the Hilbert series of $\BC^2/\BZ_2 \times \BC^2$ given by the discrete Molien formula (see \eg~\cite{Benvenuti:2006qr}):
\bea
g_{\BC^2/\BZ_2 \times \BC^2 }(t,x)
&=&\frac{1}{2} \left[ \frac{1}{\left(1-\frac{t}{x}\right) (1-t x)}+\frac{1}{\left(1+\frac{t}{x}\right) (1+t x)} \right] \times \frac{1}{\left(1-\frac{t}{x}\right) (1-t x)} \nn \\
&=& (1-t^4) \PE \left[ [1] t + [2] t^2 \right]~.
\eea
is equal to the Hilbert series \eref{HStad}.

\subsubsection{The tadpole from gluing two legs in the $T_2$ theory} \label{selfglue1}
It is clear from the skeleton diagram that the tadpole comes from gluing two legs of the $T_2$ theory.  Let us derive the corresponding gluing factor.  Starting from \eref{T2PE}, we glue the legs 1 and 2 together (\emph{i.e.} set $x_1=x_2=z$ and take $x_3=x$); we then obtain
\bea
[1;1;1] \equiv ([1]_z  [1]_z) [1]_x= ([2]_z + [0]_z) [1]_x \equiv [2;1] + [0;1]~,
\eea
where $[a,b]= [a]_z [b]_x$.
Observe that this is actually the representation in the plethystic exponential \eref{fftad}.
Hence, from \eref{fftad}, it is immediate that the gluing factor is
\bea
g_{\mathrm{glue}}(t,z,x) = 1-t^2[2;0]+t^3[0;1]~. \label{exselfglue}
\eea
Let us comment on the gluing factor as follows:
\bi
\item This process involves self-gluing. The gluing factor is different from \eref{gluingfac}.
\item Whenever the self-gluing gets involved, the gluing factor is no longer local.  As can be seen from \eref{exselfglue}, the gluing factor does not depend only on $z$, the variable associated with the two legs we glue, but it depends also on $x$, the variable associated with the third leg which is not involved in the gluing. 
\item When there is no self-gluing involved, the gluing is a local process and the gluing factor is given by \eref {gluingfac}.
\ei

\subsection{The theories with genus one and two external legs $(g=1, e=2)$}
 Below the Hilbert series of theories with genus one and two external legs are studied in detail.
 
\subsubsection{The $A_1$ theory}
\begin{figure}[htbp]
\begin{center}
\includegraphics[height=2 cm]{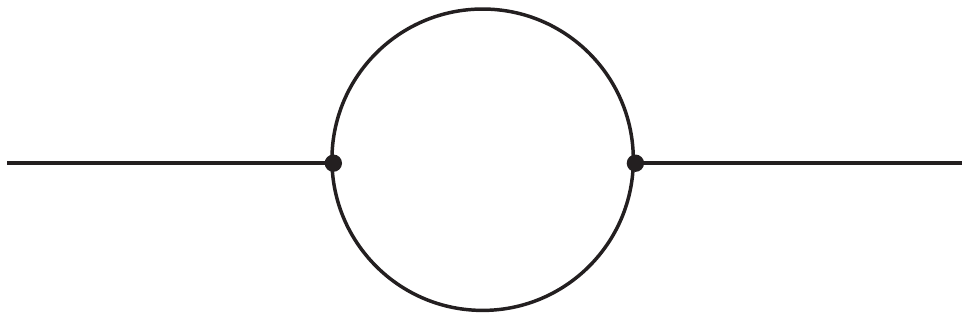}
\caption{The skeleton diagram of the $A_1$ theory.}
\label{A1skel}
\end{center}
\end{figure}

\begin{figure}[htbp]
\begin{center}
\includegraphics[height=1.5 cm]{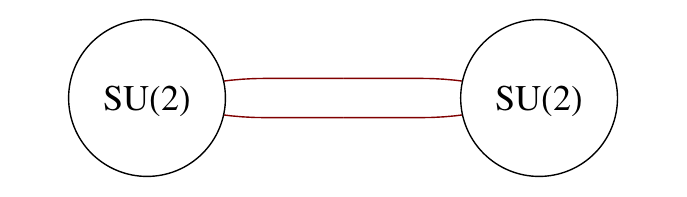}
\caption{The $\cN=2$ quiver diagram of the $A_1$ theory.}
\label{A1quiv}
\end{center}
\end{figure}

In this subsection, we focus on the theory with the $A_1$ quiver, whose skeleton diagram is depicted in \fref{A1skel}.  The two $SU(2)$ gauge groups are represented by the upper and lower arcs.  The two external legs represent the two $SU(2)$ baryonic symmetries, $SU(2)_{B_1}$ and $SU(2)_{B_2}$.  
The quiver diagram of the $A_1$ theory is given by \fref{A1quiv}.

Let $\phi_1$ and $\phi_2$ be the scalar fields in the two $\cN=2$ $SU(2)$ vector multiplets. In an $\cN=1$ notation, the superpotential can be written according to \eref{gensup} as
\bea
W &=& \cQ_{a_1 a_2 i_1} \cQ_{a'_1 a'_2 i'_1} (\phi_1^{a_1 a'_1} \epsilon^{a_2 a'_2} \epsilon^{i_1 i'_1} + \epsilon^{a_1 a'_1} \phi_2^{a_2 a'_2} \epsilon^{i_1 i'_1} ) \nn \\
&& + \tcQ_{a_1 a_2 i_2} \tcQ_{a'_1 a'_2 i'_2} (\phi_1^{a_1 a'_1} \epsilon^{a_2 a'_2} \epsilon^{i_2 i'_2} + \epsilon^{a_1 a'_1} \phi_2^{a_2 a'_2} \epsilon^{i_2 i'_2} )~,
\eea 
where for simplicity the mass terms of $\cQ$ and $\tcQ$ are set to zero.

\paragraph{The F-terms.} We first start from the F-terms of the $A_1$ theory.  Since we focus on the Kibble branch, the vacuum expectation values of $\phi_1$ and $\phi_2$ are zero.  Therefore, the non-trivial F-terms associated with the Kibble branch are the derivatives of the superpotential with respect to $\phi_1$ and $\phi_2$.  The F-terms can be written as
\bea
\mathcal{F}^1_{a_1 a'_1} = (\cQ_{a_1 a_2 i_1} \cQ_{a'_1 a'_2 i'_1} \epsilon^{i_1 i'_1} +  \tcQ_{a_1 a_2 i_2} \tcQ_{a'_1 a'_2 i'_2} \epsilon^{i_2 i'_2}) \epsilon^{a_2 a'_2} =0~, \nn \\
\mathcal{F}^2_{a_2 a'_2} = (\cQ_{a_1 a_2 i_1} \cQ_{a'_1 a'_2 i'_1} \epsilon^{i_1 i'_1}+  \tcQ_{a_1 a_2 i_2} \tcQ_{a'_1 a'_2 i'_2} \epsilon^{i_2 i'_2}) \epsilon^{a_1 a'_1} =0 ~.
\eea 

\paragraph{Dimension.} Now let us compute the dimension of the F-flat space (\ie~the space of the F-term solutions).  Since there are two nodes in the skeleton diagrams, there are $8+8 =16$ half-hypermultiplets, corresponding to $16$ complex dimensional space.  The F-terms impose 5 complex relations.  Hence, the F-flat space is $16-5=11$ complex dimensional.  Due to the $\cN=2$ supersymmetry, the D-terms also impose 5 complex relations.  Hence, the Kibble branch is $11 -5 =6$ complex dimensional, in agreement with \eref{dimmod}.

\paragraph{The Hilbert series of the F-flat space.}
This is given by
\bea
\f(t,z_1,z_2,x_1,x_2) = \mathcal{C}(t,z_1,z_2,x_1,x_2) \PE \left[ [1;1;1;0] t + [1;1;0;1] t\right]~,
\eea
where $[a;b;c;d] = [a]_{z_1} [b]_{z_2} [c]_{x_1} [d]_{x_2}$ and
\bea
\mathcal{C} &=& 1 - t^2 \left( [2;0;0;0]+[0;2;0;0] \right) + t^4 \left( [2;2;0;0] + [2;0;0;0] + [0;2;0;0] +[0;0;1;1] \right)  \nn \\
&& - t^5 \left( [1;1;1;0]+[1;1;0;1] \right) -t^6 \left( [2;2;0;0]+1 \right)+t^7 \left( [1;1;1;0]+[1;1;0;1]\right) \nn \\
&& - t^8 \left( [0;0;1;1] +1 \right)~.
\eea
Setting $z_1 = z_2 = x_1 = x_2 =1$, we obtain the unrefined Hilbert series:
\bea
\f(t,1,1,1,1) = \frac{(1+t) \left(1+4 t+5 t^2\right)}{(1-t)^{11}}~.
\eea
The pole at $t=1$ is at order 11, so the F-flat space is 11 dimensional as expected.

\paragraph{The Kibble branch Hilbert series.}  This can be obtained by integrating over the gauge fugacities:
\bea
g_{A_1}(t,x_1,x_2) = \int \ud \mu_{SU(2)}(z_1) \ud \mu_{SU(2)}(z_2) \f(t,z_1,z_2,x_1,x_2)~,
\eea
where the Haar measure of $SU(2)$ is 
\bea
\int \ud \mu_{SU(2)}(z) = \oint_{|z|=1} \frac{1-z^2}{z} \ud z~.
\eea
Evaluating this integral, one obtain a rational function of $t, x_1, x_2$ whose power series is given by
\bea
g_{A_1}(t,x_1,x_2) &=& \frac{1}{1-t^4} \sum_{n_1, n_2,m = 0}^\infty [2n_1+m; 2n_2+m] t^{2n_1+2n_2+2 m} \nn\\
&& + [2n_1+m+1; 2n_2+m+1] t^{2n_1+2n_2+2m+4}  ~. \label{A1HS}
\eea
Note that this expression is invariant under a permutation of the two external legs. The permutation group $S_2$ acts on exchanging the legs and the Hilbert series on the Kibble branch is an invariant function of this $S_2$.

The unrefined Hilbert series is
\bea
g_{A_1}(t,1,1)= \frac{\left(1+t^2\right) \left(1+3 t^2+t^4\right)}{(1-t^2)^6}~.
\eea
Note that the Kibble branch is indeed 6 complex dimensional, as expected.
The plethystic logarithm of \eref{A1HS} is given by
\bea
\PL \left[g_{A_1}(t,x_1,x_2) \right] = t^2 \left( [2;0] +[1;1] +[0;2] \right) -t^4 \left( [1;1] +2 [0;0]\right) + \ldots~.
\eea
The generators are listed in \tref{t:A1gen}.  

Since $SU(2) \times SU(2) \cong SO(4)$, it can be seen the generators transform in the 10 dimensional second rank symmetric representation\footnote{We denote the $SO(4)$ 2-dimensional spinor representation and its conjugate respectively by $[1,0]$ and $[0,1]$.  Therefore, the $SO(4)$ vector representation is $[1,1]$, and the second rank symmetric traceless representation is $[2,2]$.} (\ie~$[2,2]+[0,0]$) of $SO(4)$.

\begin{table}[htdp]
\begin{center}
\begin{tabular}{|c|c|}
\hline
Representation of the global & Generators \\
$SU(2) \times SU(2)$ & ~\\
\hline
$[2;0]$ & $M^{[2;0]}_{i_1 i'_1} = \epsilon^{a_1 a'_1} \epsilon^{a_2 a'_2} \cQ_{a_1 a_2 i_1} \cQ_{a'_1 a'_2 i'_1}$ \\
$[1;1]$ & $M^{[1;1]}_{i_1 i_2} =\epsilon^{a_1 a'_1} \epsilon^{a_2 a'_2} \cQ_{a_1 a_2 i_1} \tcQ_{a'_1 a'_2 i_2} $ \\
$[0;2]$ & $M^{[0;2]}_{i_2 i'_2} =\epsilon^{a_1 a'_1} \epsilon^{a_2 a'_2} \tcQ_{a_1 a_2 i_2} \tcQ_{a'_1 a'_2 i'_2}$ \\
\hline
\end{tabular}
\end{center}
\caption{The generators of the $A_1$ theory and the representations in which they transform.}
\label{t:A1gen}
\end{table}%

Note that the $A_1$ theory with the $U(2) \times U(2)$ gauge group is considered in \S 4.2 of \cite{Forcella:2007wk}, where two generators, namely $M^{[1;1]}_{12}$ and $M^{[1;1]}_{21}$, are set to be equal due to imposing the F-term relation for the $U(1)$ part. However, such a relation is not imposed in our analysis.

\subsubsection{The stickman model}
\begin{figure}[htbp]
\begin{center}
\includegraphics[height=3cm]{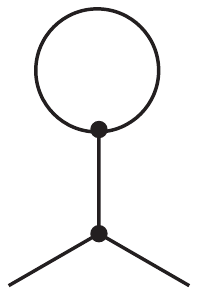}
\caption{The stickman model}
\label{man}
\end{center}
\end{figure}

The skeleton diagram of the stickman model is depicted in \fref{man}.  This model can be obtained from gluing the tadpole theory with the $T_2$ theory along the external legs.

The Hilbert series can be obtained as follows:
\bea
g_{\text{man}}(t,x_1,x_2) = \int \ud \mu_{SU(2)} (z)  g_{T_2} (t,x_1,x_2,z) g_{\mathrm{glue}} (t,z) g_{\mathrm{tadpole}} (t,z)~,
\eea
where $g_{T_2}$ is given by \eref{T2}, $g_{\mathrm{tadpole}}$ is given by \eref{HStad}, and the gluing factor $g_{\mathrm{glue}}$ is given by \eref{gluingfac}.  In order to evaluate this integral, we use the the identities \eref{iden1} and follow \eref{contr1} and \eref{contr2}.  The result is
\bea
g_{\text{man}}(t,x_1,x_2) &=& \frac{1}{1-t^4} \sum_{n_1, n_2,m = 0}^\infty [2n_1+m; 2n_2+m] t^{2n_1+2n_2+2 m} \nn\\
&& + [2n_1+m+1; 2n_2+m+1] t^{2n_1+2n_2+2m+4}  ~. \label{manHS}
\eea
Note that \eref{manHS} is equal to \eref{A1HS}.  This is a consistency check of the duality conjecture.

\section{Theories with zero external legs} \label{sec:zeroleg}
In this section, we focus on the theories with no external legs.  This class of theories has a number of interesting features.  Let us mention one of them as follows. From \eref{dimmod}, the Kibble branch of these theories is 2 complex dimensional, or equivalently 1 quarternionic dimensional.  Note that a non-compact hyperK\"ahler manifold with 1 quaternionic dimension is also known as the asymptoptic locally Euclidean (ALE) space.  Hence, we expect the Kibble branch of the theories with no external leg to be $\BC^2/\Gamma$, where $\Gamma$ is a finite subgroup of $SU(2)$.  Later we show that $\Gamma = \hat{D}_{g+1}$ for $g$ the genus of the skeleton diagram.

In the subsequent subsections, we discuss this class of theories in detail.

\subsection{The theories with genus two $(g=2, e=0)$}
\begin{figure}[htbp]
\begin{center}
\includegraphics[height=2cm]{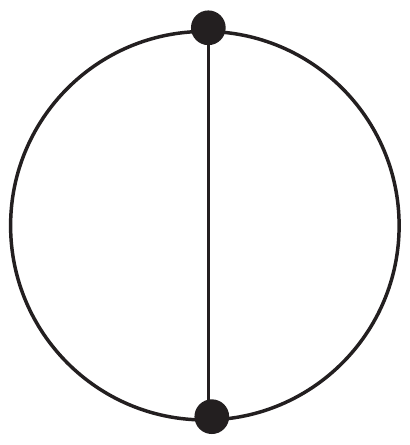}
\hspace{3cm}
\includegraphics[height=2cm]{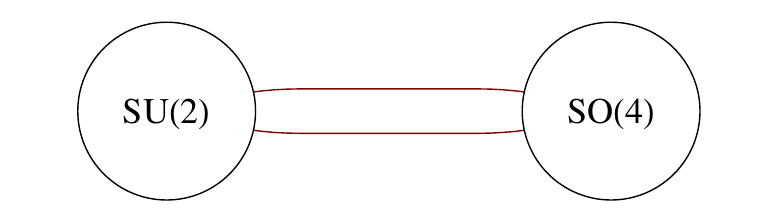}
\caption{(i) A skeleton diagram with $(g=2,e=0)$. We refer to this diagram as the {\bf Yin-Yang diagram}. (ii) The corresponding quiver digram. The $SO(4) = SU(2) \times SU(2)$ gauge group arises from the two $SU(2)$ gauge groups corresponding to the left and the right arcs in the skeleton diagram. The two lines corresponds to the 8 half-hypermultiplets (two nodes in the skeleton diagram).}
\label{g2e0a}
\end{center}
\end{figure}

\begin{figure}[htbp]
\begin{center}
\includegraphics[height=1.5cm]{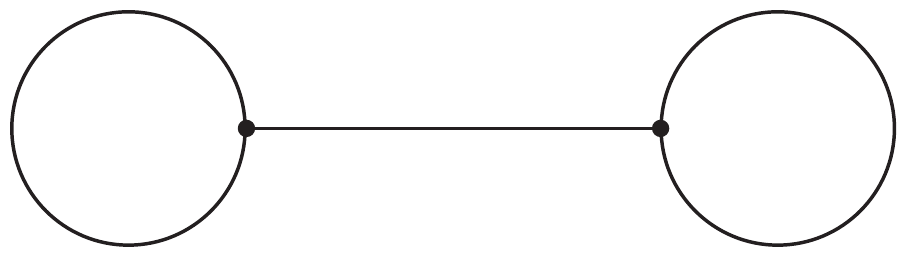}
\caption{(Dumbbell) Another skeleton diagram with $(g=2,e=0)$. }
\label{dumbbell}
\end{center}
\end{figure}

There are two skeleton diagrams corresponding to $(g=2, e=0)$.  The first one, which we will refer to as the {\bf Yin-Yang diagram}\footnote{The name comes from the Yin-Yang symbol 
\includegraphics[height=0.5cm]{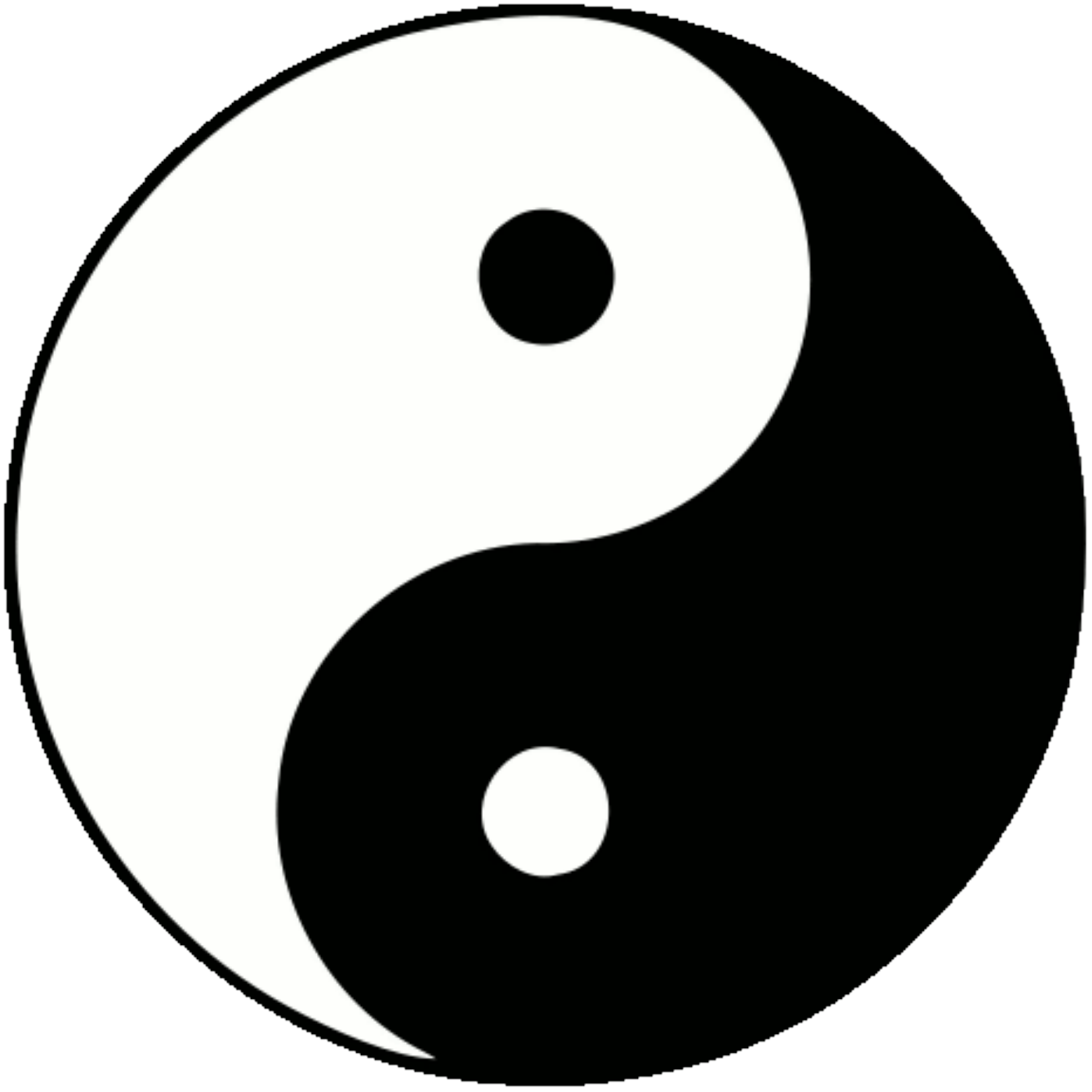}.}, is depicted in \fref{g2e0a}(i).  The corresponding quiver diagram is given in \fref{g2e0a}(ii).  The $SO(4) = SU(2) \times SU(2)$ gauge group comes from the two $SU(2)$ gauge groups corresponding to the left and the right arcs in the skeleton diagram. The two lines correspond to the 8 half-hypermultiplets (\ie~two nodes in the skeleton diagram).  The second skeleton diagram, which we will refer to as the {\bf dumbbell diagram}, is depicted in \fref{dumbbell}.  

We subsequently compute the Hilbert series of the Yin-Yang model and the dumbbell model and show that they are equal.  This again demonstrates that the Kibble branch depends only on the topology of the skeleton diagram, but not on other details of the diagram.

\subsubsection{The dumbbell model}
The skeleton of the dumbbell model is depicted in \fref{dumbbell}.  This model can be obtained by gluing the tails of two tadpoles.
Therefore, the Hilbert series of the Kibble branch of this model is
\bea
g_{D}(t) &=& \int \ud \mu_{SU(2)} (z) g_{\text{tadpole}} (t, z) g_{\text{glue}} (t,z) g_{\text{tadpole}} (t, z) \nn \\
&=& \oint_{|z|=1} \frac{\ud z}{z} (1-z^2) \frac{(1-t^4)^2 \PE \left[ 2[1]_z t + 2[2]_z t^2 \right]}{\PE \left[ [2]_z t^2 \right]}  \nn\\
&=& \frac{1-t^8}{(1-t^2)(1-t^4)^2}~. \label{HSdumb}
\eea
where the gluing factor $g_{\text{glue}} (t,z)$ is given by \eref{gfsu2} and $g_{\text{tadpole}} (t,z)$ is given by \eref{HStad}.
Observe that the Kibble branch is a two complex dimensional complete intersection.
There is one generator at order $t^2$, two generators at order $t^4$, and one relation at order $t^8$.
Note that this is the Hilbert series of $\BC^2/\hat{D}_3$ \cite{Benvenuti:2006qr}.

\subsubsection*{The generators of the moduli space}
The dumbbell model has three $SU(2)$ gauge groups: one corresponds to the left loop (denoted by $SU(2)_1$), one corresponds to the line (denoted by $SU(2)_2$), and one corresponds the right loop (denoted by $SU(2)_3$).  In an $\cN=1$ notation, the matter content is tabulated in \tref{matterD}.  The superpotential, according to \eref{gensup}, is
\bea
W &=& \cQ_{a_1 b_1 a_2} \cQ_{a'_1 b'_1 a'_2} (\phi_1^{a_1 a'_1} \epsilon^{b_1 b'_1} \epsilon^{a_2 a'_2} + \epsilon^{a_1 a'_1} \phi_1^{b_1 b'_1} \epsilon^{a_2 a'_2} + \epsilon^{a_1 a'_1} \epsilon^{b_1 b'_1} \phi_2^{a_2 a'_2} )  \nn \\
&&+\tcQ_{a_3 b_3 a_2} \tcQ_{a'_3 b'_3 a'_2} (\phi_3^{a_3 a'_3} \epsilon^{b_3 b'_3} \epsilon^{a_2 a'_2} + \epsilon^{a_3 a'_3} \phi_3^{b_3 b'_3} \epsilon^{a_2 a'_2} + \epsilon^{a_3 a'_3} \epsilon^{b_3 b'_3} \phi_2^{a_2 a'_2} )~.
\eea 
\begin{table}[htdp]
\begin{center}
\begin{tabular}{|c||c|c|c|}
\hline 
Field & Gauge   & Gauge & Gauge  \\
& $SU(2)_1$ & $SU(2)_2$ & $SU(2)_3$ \\
\hline \hline
 $\phi^{a_1 a'_1}_1$ & $[2]$ & $[0]$ & $[0]$ \\
 $\cQ_{a_1 b_1 a_2}$ & $[2]+[0]$ & $[1]$ & $[0]$ \\
\hline \hline
 $\phi^{a_2 a'_2}_2$ & $[0]$ & $[2]$ & $[0]$ \\
\hline \hline
 $\phi^{a_3 a'_3}_3$ & $[0]$ & $[0]$ & $[2]$ \\
$\tcQ_{a_3 b_3 a_2}$ & $[0]$ & $[1]$ & $[2]+[0]$ \\
\hline
\end{tabular}
\caption{The matter content in the $\cN=1$ language of the dumbbell theory.}
\label{matterD}
\end{center}
\end{table}%

Note that on the Kibble branch the VEVs of $\phi_1, \phi_2, \phi_3$ are zero. Hence, the non-trivial F-terms come from the derivatives of $\phi_1, \phi_2, \phi_3$:
\bea
\epsilon^{b_1 b'_1} \epsilon^{a_2 a'_2} (\cQ_{a_1 b_1 a_2} \cQ_{a'_1 b'_1 a'_2} +\cQ_{b_1 a_1 a_2} \cQ_{b'_1 a'_1 a'_2})  &=& 0~, \nn \\
\epsilon^{a_1 a'_1} \epsilon^{b_1 b'_1} \cQ_{a_1 b_1 a_2} \cQ_{a'_1 b'_1 a'_2} +  \epsilon^{a_3 a'_3} \epsilon^{b_3 b'_3}\tcQ_{a_3 b_3 a_2} \tcQ_{a'_3 b'_3 a'_2} &=& 0~, \nn\\
\epsilon^{b_3 b'_3} \epsilon^{a_2 a'_2} (\tcQ_{a_3 b_3 a_2} \tcQ_{a'_3 b'_3 a'_2} + \tcQ_{b_3 a_3 a_2} \tcQ_{b'_3 a'_3 a'_2} )&=& 0~. \label{FQ}
\eea
\comment{
As for the tadpole theory, we split $\cQ$ and $\tcQ$ into the trace and the traceless part as
\bea
A_\alpha &=& \cQ^a_{~a, \alpha}~, \quad \varphi^a_{~b, \alpha} = \cQ^a_{~b, \alpha} - \frac{1}{2} A_\alpha \delta^a_{~b}~, \nn \\
B_\alpha &=& \tcQ^i_{~i, \alpha}~, \quad \psi^i_{~j, \alpha} = \tcQ^i_{~j, \alpha} - \frac{1}{2} B_\alpha \delta^i_{~j}~,
\eea
The F-terms \eref{FQ} can then be written in terms of matrix relations as
\bea
[ \varphi_1, \varphi_2] = [\psi_1, \psi_2] = 0~, \label{commu} \\
U_{\alpha \beta} + V_{\alpha \beta} + A_\alpha A_\beta + B_\alpha B_\beta =0~, \label{Thet}
\eea
where
\bea
U_{\alpha \beta} = \tr(\varphi_\alpha \cdot \varphi_\beta)~, \qquad V_{\alpha \beta} = \tr(\psi_\alpha \cdot \psi_\beta)~.
\eea
The relations \eref{commu} imply that
\bea
\det U = \det V = 0~.
\eea
}
The generator at order $t^2$ is
\bea
M = \epsilon^{a_1 b_1} \epsilon^{a_2 a'_2} \epsilon^{a_3 b_3}  \cQ_{a_1 b_1 a_2}  \tcQ_{a_3 b_3 a'_2} ~.
\label{Mdumb}
\eea
The generators at order $t^4$ are
\bea
\cB_1 &=&  \epsilon^{a_1 b_1}  \epsilon^{a'_1 b'_1}  \epsilon^{a_2 b_2}  \epsilon^{a'_2 b'_2}  \cQ_{a_1 b_1 a_2} U_{b_2 b'_2}\cQ_{a'_1 b'_1 a'_2}~, \nn \\
\cB_2 &=&  \epsilon^{a_1 b_1}  \epsilon^{a'_3 b'_3} \epsilon^{a_2 b_2}  \epsilon^{a'_2 b'_2} \cQ_{a_1 b_1 a_2} U_{b_2 b'_2}\tcQ_{a'_3 b'_3 a'_2}~.
\eea
where 
\bea
U_{a_2 a'_2} = \epsilon^{a_1 a'_1} \epsilon^{b_1 b'_1}  \cQ_{a_1 b_1 a_2} \cQ_{a'_1 b'_1 a'_2}~. \label{defU}
\eea
Note that by the second F-terms in \eref{FQ}, it follows that
\bea
\epsilon^{a_3 a'_3} \epsilon^{b_3 b'_3}  \tcQ_{a_3 b_3 a_2} \tcQ_{a'_3 b'_3 a'_2} = - U_{a_2 a'_2}~. \label{minusU}
\eea
Other order $4$ operators can be expressed in terms of $M$, $\cB_1$ and $\cB_2$ as follows:
\bi
\item 
Using \eref{defU} and the first F-terms in \eref{FQ}, we obtain
\bea
\det U = \frac{1}{2} \epsilon^{a_2 b_2} \epsilon^{a'_2 b'_2} U_{a_2 a'_2} U_{b_2 b'_2} = \frac{1}{2} \cB_1. \label{detU}
\eea
\item Consider the operator
\bea
\widetilde{\cB}_1 &=&  \epsilon^{a_3 b_3}  \epsilon^{a'_3 b'_3}  \epsilon^{a_2 b_2}  \epsilon^{a'_2 b'_2}  \tcQ_{a_3 b_3 a_2} U_{b_2 b'_2}\tcQ_{a'_3 b'_3 a'_2}~.
\eea
From \eref{minusU} and \eref{detU}, we have
\bea
\widetilde{\cB}_1 = - \epsilon^{a_2 b_2}  \epsilon^{a'_2 b'_2}   U_{a_2 a'_2} U_{b_2 b'_2} = -{\cB}_1~. \label{tB1}
\eea

\ei

\subsubsection*{The relation between the generators}
In order to obtain the relation, we start from the following identity which is true for any symmetric matrix $U_{ab}$.
\bea
 (\epsilon^{ab} A_a B_b)^2 \det U + (\epsilon^{aa'} \epsilon^{bb'}A_a U_{a'b'} B_b)^2- (\epsilon^{aa'} \epsilon^{bb'}A_a U_{a'b'} A_b) (\epsilon^{cc'} \epsilon^{dd'}B_c U_{c'd'} B_d) = 0~. \nn \\ \label{idendumb}
\eea
Taking $U$ to be as in \eref{defU} and taking
\bea
A_{a_2} = \epsilon^{a_1 b_1} \cQ_{a_1 b_1 a_2}, \qquad B_{a_2} =  \epsilon^{a_3 b_3} \cQ_{a_3 b_3 a_2}~,
\eea
we obtain
\bea
M^2 \det U + \cB_2^2 - \cB_1 \widetilde{\cB}_1 = 0~.
\eea
Substituting in it the identities \eref{detU} and \eref{tB1}, we obtain
\bea
2 M^2 \cB_1 + \cB_1^2 + \cB_2^2 = 0~. \label{reldumb}
\eea
Note that this is indeed the relation of $\BC^2/\hat{D}_3$.\footnote{Note that the relation for $\BC^2/\hat{D}_3$ can be written as $u^2+v^2 w =w^2$ (see \eg~\cite{Benvenuti:2006qr}), where in this case $u = i \cB_2, v = \sqrt{2} M, w = -\cB_1$.} 

\comment{
\noindent {\bf The proof of \eref{idendumb}.}  By the Cayley-Hamilton theorem, any $2 \times 2$ matrix $Y$ satisfies its own characteristic equation:
\bea
(Y^2)_{a b} - (\tr Y) Y_{ab} + (\det Y) \epsilon_{ab} = 0~, \label{CH}
\eea
where $(Y^2)_{a b}= \epsilon^{a' b'} Y_{a a'} Y_{b' b}$ and $\tr Y = \epsilon^{ab} Y_{ab}$.
Take
\bea
Y_{ab} = \epsilon^{cc'} \epsilon^{dd'} (-U_{ac} A_{c'} B_{d'} V_{d b} + U_{ac} B_{c'} A_{d'} V_{db})~.
\eea
Multiplying $(A^\alpha B^\beta-B^\alpha A^\beta) $ to \eref{CH}, we have
\bea
(Y^2)_{\alpha \beta} (A^\alpha B^\beta-B^\alpha A^\beta) - (\tr Y) Y_{\alpha \beta} (A^\alpha B^\beta-B^\alpha A^\beta) + 2M (\det Y) = 0~. \label{in1}
\eea
Let us consider
\bea
Y_{\alpha \beta} = -U_{\alpha \lambda} A^\lambda B^\sigma V_{\sigma \beta} + U_{\alpha \lambda} B^\lambda A^\sigma V_{\sigma \beta}~,
\eea
Multiplying $(A^\alpha B^\beta-B^\alpha A^\beta) $ to \eref{CH}, we have
\bea
(Y^2)_{\alpha \beta} (A^\alpha B^\beta-B^\alpha A^\beta) - (\tr Y) Y_{\alpha \beta} (A^\alpha B^\beta-B^\alpha A^\beta) + 2M (\det Y) = 0~. \label{in1}
\eea
Note the following identities:
\bea
\tr Y &=&  -MU^\alpha_{~\beta} V^\beta_{~\alpha} = - M \tr(UV)~, \nn \\
\det Y &=& M^2 \det (UV)~, \nn \\
(Y^2)_{\alpha \beta} (A^\alpha B^\beta-B^\alpha A^\beta) &=& M^2 \tr((UV)^2)~, \nn \\
Y_{\alpha \beta} (A^\alpha B^\beta-B^\alpha A^\beta) &=& 2(A^\alpha U_{\alpha \beta} B^\beta)(A^\lambda V_{\lambda \sigma} B^\sigma)-(A^\alpha U_{\alpha \beta} A^\beta)(B^\lambda V_{\lambda \sigma} B^\sigma)  \nn \\
&& -(B^\alpha U_{\alpha \beta} B^\beta)(A^\lambda V_{\lambda \sigma} A^\sigma)~, \label{in2}
\eea
We need another identity which can be obtained from taking $Y$ in \eref{CHm} to be $UV$ and taking the trace:
\bea
\tr ((UV)^2) -  \left( \tr(UV) \right)^2 + 2 \det(UV) = 0~. \label{in3}
\eea
Using \eref{in1}, \eref{in2} and \eref{in3}, we arrive at \eref{idendumb} as required.}

\subsubsection{The Yin-Yang model}
\begin{figure}[htbp]
\begin{center}
\includegraphics[height=2 cm]{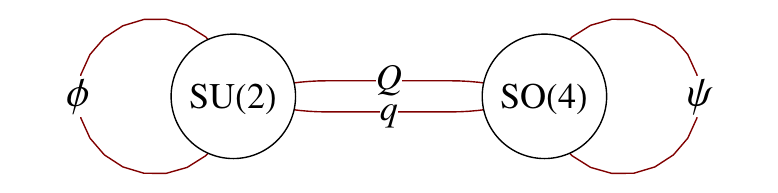}
\caption{The $\cN=1$ quiver digram of the Yin-Yang model. The superpotential is given by $W = (Q\cdot \psi \cdot Q - Q \cdot \varphi \cdot Q) +(q\cdot \psi \cdot q - q \cdot \varphi \cdot q)$.}
\label{N1YY}
\end{center}
\end{figure}

In this subsection, we compute the Hilbert series of the Yin-Yang model from the $\cN=1$ quiver diagram depicted in \fref{N1YY}.  
The bi-fundamental hypermultiplets are denoted by $Q^a_i$ and $q^a_i$, where we use $a,b,c = 1,2$ to denote the $SU(2)$ indices and $i,j,k=1,\ldots,4$ to denote the $SO(4)$ indices.  The adjoint fields in $SU(2)$ and $SO(4)$ are denoted respectively by $\varphi$ and $\psi$.  The superpotential is
\bea
W = \left( \epsilon_{ab} Q^a_i \psi^{ij} Q^b_j - \delta^{ij} Q^a_i \varphi_{ab} Q^b_j \right) +\left( \epsilon_{ab} q^a_i \psi^{ij} q^b_j - \delta^{ij} q^a_i \varphi_{ab} q^b_j \right) ~, \label{supYY}
\eea
where the $SU(2)$ indices are raised and lowered using the epsilon symbol and the $SO(4)$ indices are raised and lowered using Kronecker's delta.

\paragraph{The F-flat space.} Since we focus on the Kibble branch, the vacuum expectation values of $\varphi$ and $\psi$ are zero.  Therefore, the non-trivial F-terms associated with the Kibble branch are the derivatives of the superpotential with respect to $\varphi$ and $\psi$:
\bea
(\cF_\varphi)^{ab} &=& \delta^{ij} ( Q^a_i Q^b_j +  q^a_i q^b_j ) =0~, \nn \\
(\cF_\psi)_{ij} &=& \epsilon_{ab} ( Q^a_i  Q^b_j + q^a_i  q^b_j )=0~.
\eea
The F-flat space is 9 complex dimensional.  The fully refined Hilbert series (with the $SU(2)$ gauge fugacity $z$ and $SO(4)$ gauge fugacities $w_1, w_2$) is too long to be reported here.  Setting all of the gauge fugacities to unity, we obtain the unrefined Hilbert series
\bea
\f(t, z=1, w_1 = w_2 =1) = \frac{1+7 t+19 t^2+21 t^3+7 t^4+t^5}{(1-t)^9}~.
\eea

\paragraph{The Kibble branch Hilbert series.} This can be obtained as follows.
\bea
g_{YY} (t) = \int \ud \mu_{SU(2)}(z) \ud \mu_{SO(4)} (w_1,w_2)~\f(t,z,w_1,w_2)~,
\eea
where
\bea
\int \mu_{SO(4)} (w_1,w_2) = \oint_{|w_1|=1}  \frac{\ud w_1}{w_1} \oint_{|w_2|=1}  \frac{\ud w_2}{w_2} \left(1-\frac{w_1}{w_2} \right) (1- w_1 w_2)~.
\eea
The result of the integrations is
\bea
g_{YY} (t) = \frac{1-t^8}{(1-t^2)(1-t^4)^2} = \frac{1+t^4}{\left(1-t^2\right)^2 \left(1+t^2\right)}~. \label{HSYY}
\eea
Observe that this Hilbert series is identical to that of the dumbbell model \eref{HSdumb}.

\subsubsection{The Ying-Yang model from gluing the two legs of the $A_1$ theory}
The skeleton diagram in \fref{g2e0a} of the Yin-Yang model can be obtained by gluing the two external legs of the $A_1$ theory, 
whose skeleton diagram is depicted in \fref{A1skel}.  Note that since this gluing process involves a self-gluing, the gluing factor does not take its 
canonical form \eref{gluingfac}.  Rather, we propose that for the equation
\bea
g_{YY}(t) = \int \ud \mu_{SU(2)} (z) g_{\mathrm{glue}} (t,z) g_{A_1} (t,z,z)
\eea
with $g_{YY}(t)$ and $g_{A_1}(t,x,y)$ given respectively by \eref{HSYY} and \eref{A1HS}, a solution for $g_{\mathrm{glue}} (t,z)$ is
\bea
g_{\mathrm{glue}} (t,z) = 1- [2]_z t^2 + \frac{2 t^4 (1+t^2)}{1+t^4}~.
\eea
This solution can be verified using the following identities:
\bea
\int \ud \mu(z) \sum_{n_1, n_2, m}   [2n_1+m]_z [2n_2+m]_z~  t^{2n_1+2n_2+\chi m} &=& \frac{1}{\left(1-t^4\right) \left(1-t^{\chi }\right)} \nn \\
\int \ud \mu(z) \sum_{n_1, n_2, m}   [2n_1+m+1]_z [2n_2+m+1]_z ~   t^{2n_1+2n_2+\chi m +\chi+2} &=& \frac{t^{2+\chi }}{\left(1-t^4\right) \left(1-t^{\chi }\right)} \nn \\
\int \ud \mu(z) \sum_{n_1, n_2, m}   [2]_z  [2n_1+m]_z [2n_2+m]_z ~   t^{2n_1+2n_2+\chi m} &=& \frac{2 t^2+t^4+t^{\chi }-t^{4+\chi }}{\left(1-t^4\right) \left(1-t^{\chi }\right)}  \nn \\
\int \ud \mu(z) \sum_{n_1, n_2, m}   [2]_z [2n_1+m+1]_z [2n_2+m+1]_z ~   t^{2n_1+2n_2+\chi m +\chi+2} &=& \frac{t^{2+\chi } \left(1+2 t^2\right)}{\left(1-t^4\right) \left(1-t^{\chi }\right)}~, \nn \\ 
\eea 
Indeed, for $\chi=2$, we obtain the Hilbert series for the Ying-Yang model,
\bea
g_{YY}(t) &=& \frac{1}{1-t^4} \left[ \left( 1+\frac{2 t^4 (1+t^2)}{1+t^4} \right)\frac{1+t^4}{\left(1-t^2\right) \left(1-t^4\right)} \right. \nn\\
&& \qquad \qquad  \left. - t^2 \left(\frac{t^4 \left(1+2 t^2\right)+3 t^2+t^4-t^6}{\left(1-t^2\right) \left(1-t^4\right)} \right) \right]  \nn \\
&=& \frac{1+t^4}{\left(1-t^2\right)^2 \left(1+t^2\right)}~,
\eea
as expected.

\comment{
\paragraph{Generators and relations.} The generators at order $t^2$ are
\bea
M= \epsilon_{ab} Q^a_i q^b_i~.
\eea
The generators at order $t^4$ are
\bea
B_1= 
\eea

The generators at order $t^4$ are
\bea
B_1 &=& \epsilon^{a_1 a_2 a_3 a_4} \epsilon_{i_1 i_2}\epsilon_{i_3 i_4} Q^{i_1}_{a_1} Q^{i_2}_{a_2}q^{i_3}_{a_3} q^{i_4}_{a_4}~, \nn \\
B_2 &=& \epsilon^{a_1 a_2 a_3 a_4} \epsilon_{i_1 i_2}\epsilon_{i_3 i_4} Q^{i_1}_{a_1} q^{i_2}_{a_2}Q^{i_3}_{a_3} q^{i_4}_{a_4}~.
\eea
\todo The relation is 
\bea
M^2 B_1 - B_1^2- B_2^2 = 0~.
\eea
}

\subsection{The theories with genus three $(g=3, e=0)$}
\begin{figure}[htbp]
\begin{center}
\includegraphics[height=1.5cm]{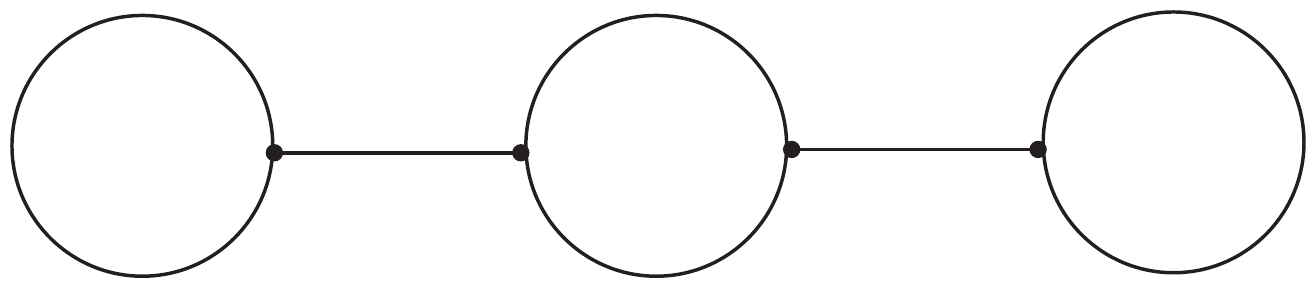}
\caption{The three-loop linear model  with zero external legs (TLMZ)}
\label{TLM}
\end{center}
\end{figure}

\begin{figure}[htbp]
\begin{center}
\includegraphics[height=2cm, angle=45]{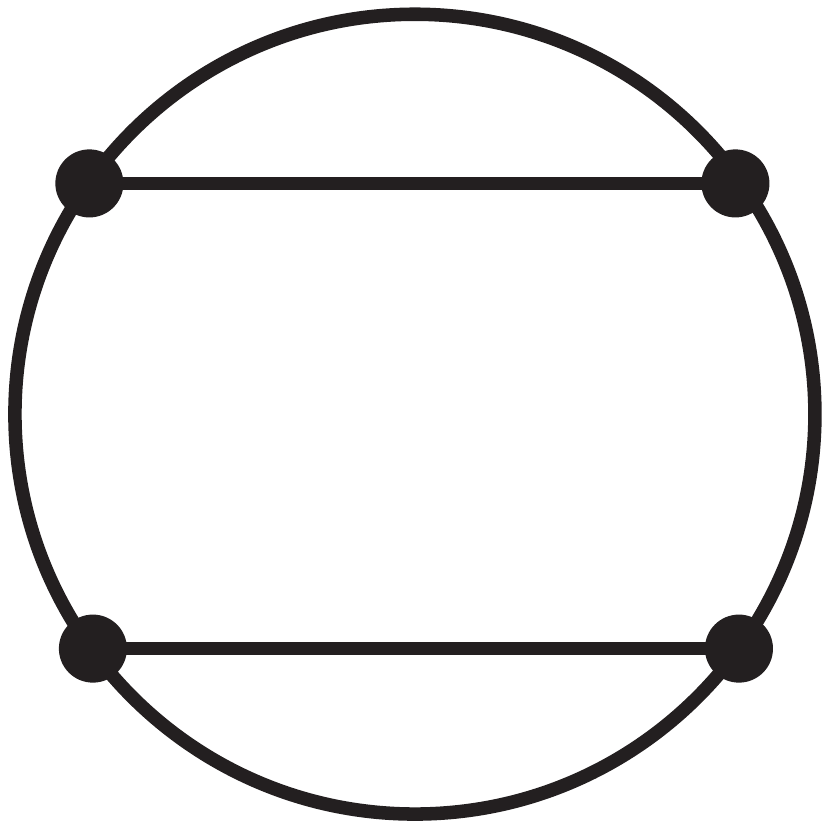}
\hspace{3cm}
\includegraphics[height=3cm, angle=-1]{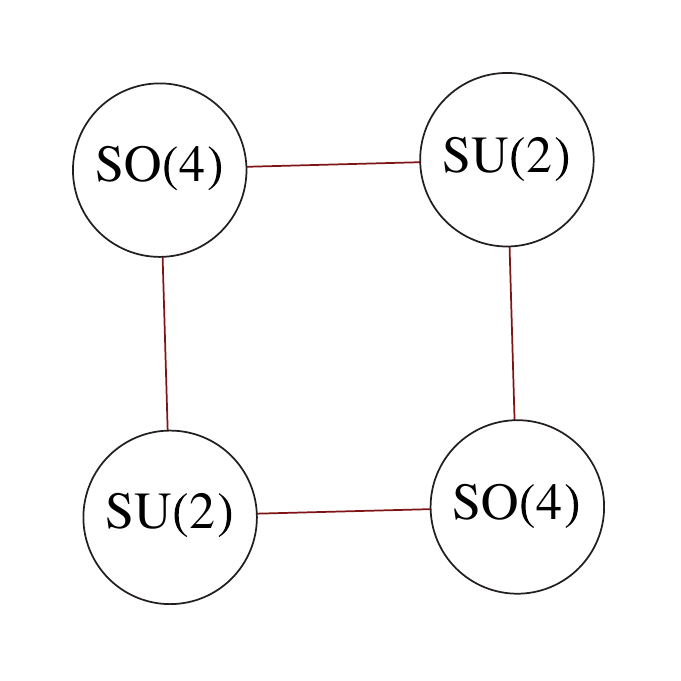}
\caption{The tablet model. {\bf Left:} The skeleton diagram. {\bf Right:} The $\cN=2$ quiver diagram. Note that each $SO(4) \cong SU(2) \times SU(2)$ gauge symmetry arises from the $SU(2)$ corresponding to the chord and the $SU(2)$ corresponding to the arc sharing the same endpoints with the chord.}
\label{tablet}
\end{center}
\end{figure}

\begin{figure}[htbp]
\begin{center}
\includegraphics[height=2cm]{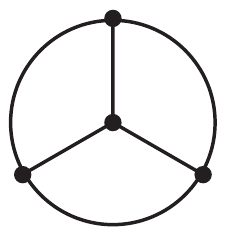}
\caption{The Mercedes-Benz model}
\label{MB}
\end{center}
\end{figure}

There are three phases of theories with genus 3 and zero external legs. Their skeleton diagrams are depicted in \fref{TLM}, \fref{tablet} and \fref{MB}. 

\subsubsection{The three-loop linear model}
In this subsection, we derive the Hilbert series of the TLMZ using the gluing technique.  

\begin{figure}[htbp]
\begin{center}
\includegraphics[height=1.5cm, angle=180]{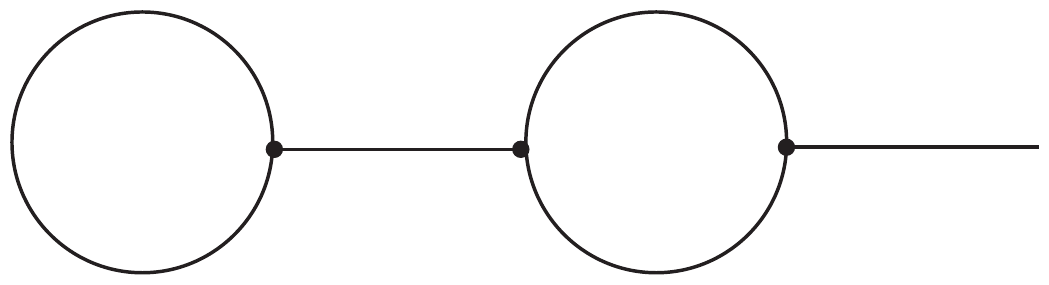}
\caption{The two loop linear model with one external leg.}
\label{twoloopslinear}
\end{center}
\end{figure}

Let us first consider the Hilbert series of the two loop linear model with one external leg depicted in \fref{twoloopslinear}.  This is given by
\bea
g_{(g=2,e=1)}(t) = \int \ud \mu_{SU(2)}(z) g_{A_1} (t,z,x)g_{\mathrm{glue}} (t,z) g_{\mathrm{tadpole}} (t,z) ~,
\eea
where $g_{\mathrm{tadpole}}$ is given by \eref{HStad} and $g_{A_1}$ is given by \eref{A1HS}.

We can evaluate this integral by using the following identities (which are a generalisation of \eref{iden1}): 
\bea
&& \int \ud \mu(z) \sum_{n_3, m, n'_3, m'} g_{\mathrm{glue}}(t,z)  [2n_3+m]_z [2n'_3+m']_z~  t^{2n_3+\chi_1 m+2n'_3+\chi_2 m'} = 
\frac{\left(1-t^2\right) \left(1+t^2-t^{2+\chi _1+\chi _2}\right)}{1-t^{\chi _1+\chi _2}}  \nn \\
&& \int \ud \mu(z) \sum_{n_3, m, n'_3, m'} g_{\mathrm{glue}}(t,z)  [2n_3+m]_z [2n'_3+m'+1]_z t^{2n_3+\chi_1 m+2n'_3+ \chi_2 m'} = \frac{t^{\chi _1} \left(1-t^2\right)}{1-t^{\chi _1+\chi _2}}   \nn \\
&& \int \ud \mu(z) \sum_{n_3, m, n'_3, m'} g_{\mathrm{glue}}(t,z) [2n_3+m+1]_z [2n'_3+m']_z t^{2n_3+\chi_1 m+2n'_3+ \chi_2 m'} = \frac{t^{\chi _2} \left(1-t^2\right)}{1-t^{\chi _1+\chi _2}}  \nn \\
&& \int \ud \mu(z) \sum_{n_3, m, n'_3, m'} g_{\mathrm{glue}}(t,z)  [2n_3+m+1]_z [2n'_3+m'+1]_z t^{2n_3+\chi_1 m+2n'_3+ \chi_2 m'} = \frac{1-t^2}{1-t^{\chi _1+\chi _2}}~. \nn \\
\label{geniden}
\eea 
Observe that for $\chi_1 = \chi_2 = 1$, we obtain the identities \eref{iden1}.  

For our problem, the $A_1$ theory has $\chi_1 =2$ and the tadpole theory has $\chi_2 =1$.  
The first and the second identities of \eref{geniden}  contribute to
\bea
&& \frac{1}{1-t^4} \left[ \frac{\left(1-t^2\right) \left(1+t^2-t^{2+\chi_1+\chi_2}\right)}{1-t^{\chi_1+\chi_2}}  + \frac{t^{\chi_2+2} \cdot t^{\chi_1} \left(1-t^2\right)}{1-t^{\chi_1+\chi_2}} \right] \nn \\
&& = \frac{1}{1-t^{\chi_1+\chi_2}} = \sum_{m=0}^\infty t^{({\chi_1+\chi_2})m}~. \label{12con}
\eea
The third and the fourth identities of \eref{geniden} contribute to
\bea
&& \frac{1}{1-t^4} \left[ \frac{t^{\chi_1+2} \cdot t^{\chi_2} \left(1-t^2\right)}{1-t^{\chi_1+\chi_2}} + \frac{t^{\chi_1+\chi_2+4} \cdot (1-t^2)}{1-t^{\chi_1+\chi_2}} \right] \nn\\
&& = \frac{t^{\chi_1+\chi_2+2}}{1-t^{\chi_1+\chi_2}} = \sum_{m=0}^\infty t^{(\chi_1+\chi_2)m+(\chi_1+\chi_2+2)}~. \label{34con}
\eea
Putting $\chi_1=2, \chi_2=1$, we obtain the Hilbert series for the two loop linear model with one external leg as
\bea
g_{(g=2, e=1)}(t,x)= \frac{1}{1-t^4} \sum_{n_1, n_2,m = 0}^\infty [2n_1+m] t^{2n_1+ 3m} + [2n_1+m+1] t^{2n_1+2n_2+3m+5}~. \nn \\
\eea

Now we compute the Hilbert series of the TLMZ (\fref{TLM}).  This is given by
\bea
g_{\text{TLM}}(t) = \int \ud \mu_{SU(2)}(z) g_{(g=2, e=1)}(t,z) g_{\mathrm{glue}}(t,z) g_{\mathrm{tadpole}} (t,z)~.
\eea
We use the identities \eref{geniden} with $\chi_1 = 3$ for the $(g=2, e=1)$ theory and $\chi_2 = 1$ for the tadpole theory.
Following \eref{12con} and \eref{34con}, we obtain the Hilbert series of the TLM as
\bea
g_{\text{TLM}}(t) = \frac{1}{1-t^4} \left( \frac{1}{1-t^{4}}+ \frac{t^6}{1-t^{4}} \right) = \frac{1+t^6}{(1-t^4)^2} = \frac{1-t^{12}}{(1-t^4)^2(1-t^6)}~.
\eea
Observe that the Kibble branch is a two complex dimensional complete intersection.
There is one generator at order $t^6$, two generators at order $t^4$, and one relation at order $t^{12}$.
Note that this is the Hilbert series of $\BC^2/\hat{D}_4$ \cite{Benvenuti:2006qr}.

\subsubsection*{The generators of the moduli space}  
\begin{figure}[htbp]
\begin{center}
\includegraphics[height=1.5cm]{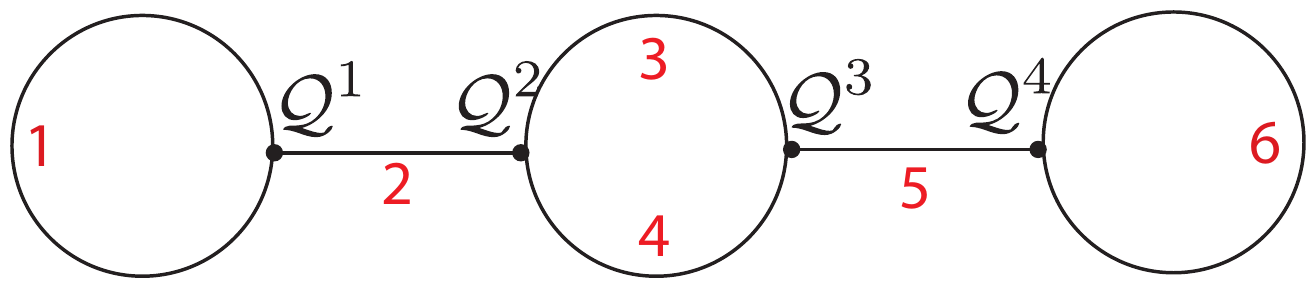}
\caption{The three-loop linear model.  The nodes are labelled by $\cQ^1, \ldots, \cQ^4$ and the gauge groups are labelled in red.}
\label{TLMwl}
\end{center}
\end{figure}
Let us label the nodes and the gauge groups according to \fref{TLMwl}.
The first generator at order 4 is
\bea
M = \epsilon^{a_1 b_1} \epsilon^{a_2 b_2} \epsilon^{a_3 b_3} \epsilon^{a_4 b_4} \epsilon^{a_5 b_5} \epsilon^{a_6 b_6}  \cQ^1_{a_1 b_1 a_2}  \cQ^2_{b_2 a_3 a_4}  \cQ^3_{b_3 b_4 a_5} \cQ^4_{b_5 a_6 b_6} ~.
\eea
Another generator at order 4 is 
\bea
\cB_1 &=&  \epsilon^{a_1 b_1}  \epsilon^{a'_1 b'_1}  \epsilon^{a_2 b_2}  \epsilon^{a'_2 b'_2}  \cQ^1_{a_1 b_1 a_2} U_{b_2 b'_2}\cQ^1_{a'_1 b'_1 a'_2}~,
\eea
where 
\bea
U_{a_2 a'_2} = \epsilon^{a_1 a'_1} \epsilon^{b_1 b'_1} \cQ^1_{a_1 b_1 a_2} \cQ^1_{a'_1 b'_1 a'_2}~.
\eea
The generator at order 6 is
\bea
\cB_2 &=& \epsilon^{a_1 b_1}  \epsilon^{a_2 b_2}  \epsilon^{a'_2 b'_2}  \epsilon^{a_3 b_3} \epsilon^{a_4 b_4} \epsilon^{a_5 b_5} \epsilon^{a_6 b_6} \cQ^1_{a_1 b_1 a_2} U_{b_2 b'_2} \cQ^2_{a'_2 a_3 a_4} \cQ^3_{b_3 b_4 a_5} \cQ^4_{b_5 a_6 b_6}~.
\eea


\section{The general formula for any genus and any external leg} \label{sec:main}
As we have seen from several example above, we claim that the Hilbert series for a theory with genus $g$ and $e$ external legs is
\bea
 g_{(g,e)}(t,x_1, \ldots, x_e) &=& \frac{1}{1-t^4} \sum_{n_1=0}^\infty  \cdots \sum_{n_e=0}^\infty \sum_{m =0}^\infty \left( \left[ 2n_1+m, \ldots, 2n_e+m\right] t^{2n_1+\ldots+2n_e+\chi m} \right. \nn \\
&& \left. + \left[ 2n_1+m+1, \ldots, 2n_e+m+1\right] t^{2n_1+\ldots+2n_e+\chi m+\chi+2} \right)~, \label{genge}
\eea
where $\chi = 2g-2+e$.

Observe that this expression is invariant under any permutation of the $e$ external legs. The permutation group $S_e$ acts on exchanging the legs and the Hilbert series on the Kibble branch is an invariant function of this $S_e$.

We prove this formula by induction in \sref{sec:proof}.  Below we discuss interesting special cases of $e=0$ and $e=1$.

\subsection{Special case: $e=0$}  
The formula \eref{genge} reduces to
\bea
g_{(g,e=0)} (t) = \frac{1-t^{4g}}{\left(1-t^4\right) \left(1-t^{2g-2} \right) \left(1-t^{2g} \right)} = \frac{1+t^{2g}}{\left(1-t^4\right) \left(1-t^{2g-2} \right)}  ~.
\eea
This Hilbert series indicates that the Kibble branch of a theory with genus $g$ and zero external legs is a two complex dimensional complete intersection.  This space is isomorphic to $\BC^2/\hat{D}_{g+1}$ \cite{Benvenuti:2006qr}.  There are generators at orders $4$, $2g-2$ and $2g$ subject to one relation at order $4g$.

\begin{figure}[htbp]
\begin{center}
\includegraphics[height=1cm]{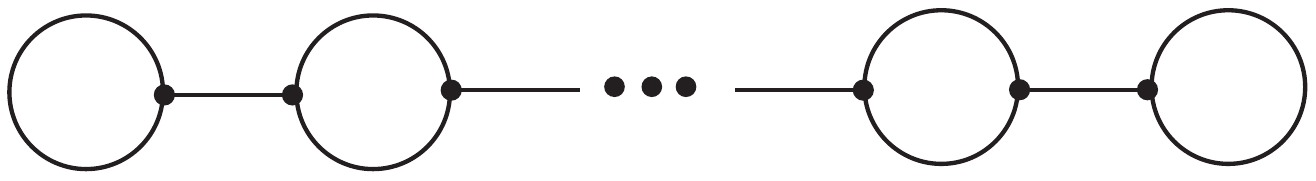}
\caption{The $g$-loop linear model with no legs}
\label{gloopnoleg}
\end{center}
\end{figure}

One can write down the generators explicitly as follows.  Consider the $g$-loop linear model with no legs depicted in \fref{gloopnoleg}.  Let us denote the nodes by $\cQ^1, \ldots, \cQ^{2g-2}$ from left to right.

The generator at order $t^{2g-2}$ can be written as
\bea
M = \epsilon^{a_1 b_1} \epsilon^{a_2 b_2} \epsilon^{a_3 b_3} \ldots \epsilon^{a_{3g-3} b_{3g-3
}}  \cQ^1_{a_1 b_1 a_2}  \cQ^2_{b_2 a_3 a_4} \ldots \cQ^{2g-2}_{b_{3g-4} a_{3g-3} b_{3g-3}} ~.
\eea
Observe that $M$ is a product of all $\cQ$'s.

The generator at order $t^4$ can be written as
\bea
\cB_1 &=&  \epsilon^{a_1 b_1}  \epsilon^{a'_1 b'_1}  \epsilon^{a_2 b_2}  \epsilon^{a'_2 b'_2}  \cQ^1_{a_1 b_1 a_2} U_{b_2 b'_2}\cQ^1_{a'_1 b'_1 a'_2}~,
\eea
where 
\bea
U_{a_2 a'_2} = \epsilon^{a_1 a'_1} \epsilon^{b_1 b'_1} \cQ^1_{a_1 b_1 a_2} \cQ^1_{a'_1 b'_1 a'_2}~.
\eea
Observe that $\cB_1$ involves in only the leftmost node $\cQ_1$.

The generator at order $t^{2g}$ can be written as
\bea
\cB_2 &=& \epsilon^{a_1 b_1}  \epsilon^{a_2 b_2}  \epsilon^{a'_2 b'_2}  \epsilon^{a_3 b_3} \ldots \epsilon^{a_{3g-3} b_{3g-3}} \cQ^1_{a_1 b_1 a_2} U_{b_2 b'_2} \cQ^2_{a'_2 a_3 a_4} \ldots \cQ^{2g-2}_{b_{3g-4} a_{3g-3} b_{3g-3}}~.
\eea

The relation at order $t^{4g}$ is given by
\bea
2 M^2 \cB_1 + \cB_1^{g} + \cB_2^2 = 0~.
\eea

\subsection{Special case: $e=1$}
The formula \eref{genge} reduces to 
\bea
g_{(g,e=1)} (t,x) &=&  \frac{1}{1-t^4} \sum_{n, m=0}^\infty \left( \left[ 2n+m \right] t^{2n+\chi m} + \left[ 2n+m+1 \right] t^{2n+\chi m + \chi +2}\right) \nn \\
&=& (1-t^{2\chi+2}) \PE \left[ [2]t^2 + [1]t^{\chi} \right]~,
\eea
where in this case $\chi = 2g-1$.  Setting $x=1$, the unrefined Hilbert series is
\bea
g_{(g,e=1)} (t,1) &=& \frac{1-t^{4g}}{(1-t^2)^3(1-t^{2g-1})^2} = \frac{1+t^{2g}}{(1-t^2)^2 (1-t^{2g-1})^2}~.
\eea
The Hilbert series indicates that the Kibble branch is a four complex dimensional complete intersection.   The generators at order $2$ transform in the $SU(2)$ representation $[2]$ and the generators at order $\chi = 2g-1$ transform in the representation $[1]$.  There is one relation at order $2\chi+2 = 4g$.

\begin{figure}[htbp]
\begin{center}
\includegraphics[height=1cm]{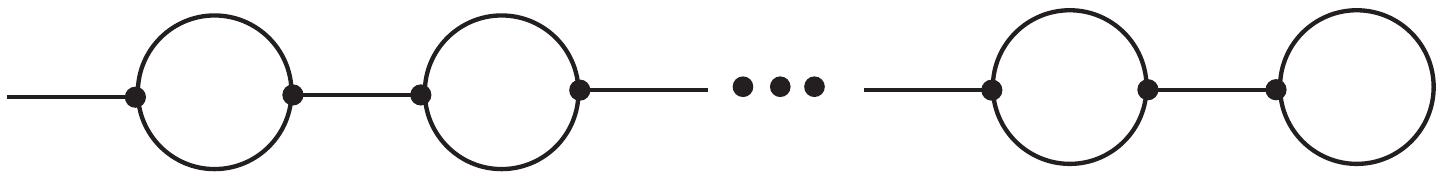}
\caption{The $g$-loop linear model with one external leg}
\label{glooponeleg}
\end{center}
\end{figure}

One can write down the generators explicitly as follows.  Consider the $g$-loop linear model with one leg depicted in \fref{glooponeleg}.  Let us denote the nodes by $\cQ^1, \ldots, \cQ^{2g-1}$ from left to right.

The generators at order $t^{2}$ can be written as
\bea
M_{i_1 i'_1} = \epsilon^{a a'} \epsilon^{b b'}  \cQ^1_{a b i_1} \cQ^1_{a' b' i'_1}~.
\eea
Observe that $M$ involves in only the leftmost node $\cQ_1$.

The generators at order $t^{2g-1}$ can be written as
\bea
B_i = \epsilon^{a_1 b_1} \epsilon^{a_2 b_2} \epsilon^{a_3 b_3} \ldots \epsilon^{a_{3g-2} b_{3g-2}}  \cQ^1_{a_1 b_1 i}  \cQ^2_{a_2 b_2 a_3} Q^3_{b_3 a_4 b_4}\ldots Q^{2g-2}_{a_{3g-4} b_{3g-4} a_{3g-3}}\cQ^{2g-1}_{b_{3g-3} a_{3g-2} b_{3g-2}} ~. \nn \\
\eea
Observe that $B$ is a product of all $\cQ$'s.

The relation at order $t^{4g}$ is of the form
\bea
(\det M)^g = f(g) M_{i_1 i_2} B_{j_1} B_{j_2} \epsilon^{i_1 j_1} \epsilon^{i_2 j_2}~,
\eea
where $f(g)$ is some function of $g$.  As an example, for $g=1$, we have $f(g)=0$ and so the relation is $\det M = 0$ (c.f. \eref{reltadpole} of the tadpole model).

\subsection{The total number of generators for any $g$ and $e$.}
In this subsection, we count the generators in a theory with any given $g$ and $e$.  

Let us focus on the case in which $\chi = 2g-2+e \geq 2$. The plethystic logarithm of \eref{genge} is 
\bea
\PL \left[ g_{(g,e)}(t,x_1, \ldots, x_e) \right] &=& \left( [2;0;\dots;0] + [0;2; \dots;0] + \ldots + [0;0;\ldots;2] \right)t^2 \nn \\
&& + [1;1; \ldots;1] t^{\chi} + \ldots~.
\eea
This indicates that the generators at order $t^2$ transform in the representation $[2;0;\dots;0] + [0;2; \dots;0] + \ldots + [0;0;\ldots;2]$ of $SU(2)^e$ and the generators at order $t^\chi$ transform in the representation $[1;1; \ldots;1]$ of $SU(2)^e$.  Hence, there are $3e$ generators at order $t^2$ and $2^e$ generators at order $t^\chi$.

For $\chi = 1$, there are only two theories, namely the $T_2$ theory and the tadpole.  In the former, there is precisely one generator $\cQ_{ijk}$.  In the latter, there is also precisely one generator $M_{ij}$ given by \eref{Mtadpole}.

\subsection{The inductive proof of the general formula} \label{sec:proof}
In this subsection, we prove the formula \eref{genge} by induction.
A key assumption we make here is that theories corresponding to different graphs with the same genus and the same number of external legs possess the same Hilbert series.  This assumption is based on the conjecture that such theories are related to each other by S-duality and has been demonstrated by several examples so far.

We are arguing that
\ben
\item Eq. \eref{genge} is true for $(g=0,e=3)$ and $(g=1, e=1)$.
\item Eq. \eref{genge} for $(g,e=1)$ implies Eq. \eref{genge} for $(g+1,e=1)$.
\item Eq. \eref{genge} for $(g,e)$ implies Eq. \eref{genge} for $(g,e+1)$.
\een
After proving these steps, we establish the formula \eref{genge} for non-trivial cases. The special case $(g,e=0)$ follows immediately as discussed above.

\paragraph{Step 1.} This can easily be done.  For $(g=0,e=3)$, we consider the $T_2$ theory \eref{T2}.  For $(g=1,e=1)$, we consider the tadpole theory \eref{HStad}.  
\paragraph{Step 2.} Assume that Eq. \eref{genge} is true for $(g,e=1)$. We glue a theory with $(g=1, e=2)$ to the $(g, e=1)$ theory along the external legs.  Hence,
\bea
g_{(g+1,e=1)}(t,x) = \int \ud \mu_{SU(2)} (z)~g_{(g, e=1)}(t,z) \ggl(t,z) g_{(g=1, e=2)}(t,z,x)
\eea
To evaluate this, we use the identites \eref{geniden} with $\chi_1 = 2g-2+1 =2g-1$ and $\chi_2 = 2$, and follow \eref{12con} and \eref{34con}.
We then obtain the expression for $(g+1,e=1)$ as 
\bea
g_{(g+1,e=1)}(t,x) &=& \frac{1}{1-t^4} \sum_{n_1=0}^\infty \sum_{m =0}^\infty  \left[ 2n_1+m \right] t^{2n_1+(2g+1)m}  \nn \\
&& + \left[ 2n_1+m+1 \right] t^{2n_1+(2g+1)m+(2g+3)}~.
\eea
This is in agreement with \eref{genge} for $(g+1,e=1)$.

\paragraph{Step 3.} Now assume that Eq. \eref{genge} is true for $(g,e)$. We glue the $T_2$ theory to the the $(g,e)$ theory along the external legs. Hence,
\bea
g_{(g,e+1)}(t,x_1, \ldots x_{e+1}) = \int \ud \mu (z)~g_{(g, e)}(t,x_1, \ldots, x_{e-1},z) \ggl(t,z) g_{T_2}(t,z,x_{e},x_{e+1})~. \nn \\
\eea
To evaluate this, we use the identites \eref{geniden} with $\chi_1 = 2g-2+e $ and $\chi_2 = 1$, and follow \eref{12con} and \eref{34con}.  We then obtain
\bea
 g_{(g,e+1)}(t,x_1, \ldots, x_{e+1}) &=& \frac{1}{1-t^4} \sum_{n_1=0}^\infty  \cdots \sum_{n_{e+1}=0}^\infty \sum_{m =0}^\infty \left( \left[ 2n_1+m, \ldots, 2n_{e+1}+m\right] t^{2n_1+\ldots+2n_{e+1}+\tilde{\chi} m} \right. \nn \\
&& \left. + \left[ 2n_1+m+1, \ldots, 2n_{e+1}+m+1\right] t^{2n_1+\ldots+2n_{e+1}+\tilde{\chi} m+\tilde{\chi}+2} \right)~,
\eea
where $\tilde{\chi} = \chi_1+\chi_2 = 2g-2+(e+1)$. This is in agreement with \eref{genge} for $(g,e+1)$.

\subsection{The general formula in terms of products}
The general formula \eref{genge} can actually be rewritten in another form involving products.  In order to do so, we use the identity
\bea \label{genfunc}
f_m(t,x) \equiv \sum_{n=0}^\infty [2n+m]_x t^{2n} = (1-t^2) \left([m]_x - [m-2]_x t^2 \right) \PE \left[[2]t^2\right]~.
\eea
Then, it is immediate that
\bea \label{prodformgen}
 g_{(g,e)} (t,x_1, \ldots, x_e) = \frac{1}{1-t^4} \sum_{m=0}^\infty \left( t^{\chi m} \prod_{i=1}^e f_m(t,x_i) + t^{\chi m+\chi+2} \prod_{i=1}^e f_{m+1}(t,x_i)  \right)~.
\eea


\acknowledgments
We would like to thank Sergio Benvenuti, Giuseppe Torri, Francesco Benini, Yuji Tachikawa, Keshav Dasgupta, Alisha Wissanji, Benjamin Hoare, Tom Pugh, Stanislav Kuperstein and Yang-Hui He for useful discussions.

N.~M.~ would like to express his gratitude to University of Pennsylvania, Princeton University, McGill University, Perimeter Institute, University of California at Los Angeles, University of California at Santa Barbara and KITP, California Institute of Technology, University of California at Berkeley, as well as Alisha Wissanji, Yong Yun and her family for their kind hospitality during the completion of this paper.  He is very grateful to Aroonroj Mekareeya for his generosity in providing his laptop computer to use in this work.  Finally, he would like to thank his family for the warm encouragement and support, as well as the DPST project and the Royal Thai Government for funding his research.

\appendix
\section{The unbroken $U(1)^g$ gauge symmetry on the Kibble branch of the theory with genus $g$} \label{sec:unbrokenu1}
\subsection{A theory with genus one} \label{sec:genus1}
As we state in \sref{sec:kibble}, at a generic point on the Kibble branch the $SU(2)$ gauge symmetry is broken to $U(1)$, corresponding to the genus of the skeleton diagram.  In this subsection, we prove to this statement by showing that two of the three components of the scalar field $\phi$ in the $SU(2)$ vector multiplet become massive and the other component remains massless. 

In this subsection, it is convenient to work with $SU(2)$ adjoint indices $A,B,C=1,2,3$.  We take the generators of the $SU(2)$ group to be $T^A = \sigma^A/2$, where $\sigma^A$ are the Pauli matrices.  We note the identity
\bea
(T^A)^a_{~b} (T^B)^b_{~c} = \frac{1}{4} \delta^{AB} \delta^{a}_{~c} + \frac{1}{2} i \epsilon^{ABC} (T^C)^{a}_{~c}~.
\eea
The adjoint fields can be written as
\bea
\phi^{a}_{~a'} = \phi^A (T^A)^{a}_{~a'}~, \qquad \varphi^{a}_{~b1} = \varphi_1^A (T^A)^{a}_{~b} ~, \qquad \varphi^{a}_{~b2} = \varphi_2^A (T^A)^{a}_{~b}~,
\eea
where $\Phi^A, \varphi_1, \varphi_2$ are complex numbers.  We emphasise that $SU(2)$ fundamental indices $a,b,a',b'$ are raised and lowered using the epsilon symbol.  The superpotential \eref{suptadphi} can then be rewritten as
\bea
W = \frac{i}{2} \epsilon^{ABC} \phi^A \varphi_1^B \varphi_2^C ~.
\eea 
The equation of motion of the $F$ auxiliary field corresponding to $\varphi_1^C$ and $\varphi_2^C$ is the (minus) derivative of $W$ with respect to $\varphi_1^C$ and $\varphi_2^C$.
\bea
-F^C_{1} &=& \frac{\partial W}{\partial \varphi^C_{1}} = -\frac{i}{2} \epsilon^{ABC} \phi^A \varphi_2^B ~, \nn \\
-F^C_{2} &=& \frac{\partial W}{\partial \varphi^C_{2}} =\frac{i}{2} \epsilon^{ABC} \phi^A \varphi_1^B~.
\eea
The potential $V$ contains the terms $(F_1^C)^*F_1^C + (F_2^C)^*F_2^C$, where
\bea
(F_1^C)^*F_1^C &=& \frac{1}{4} (\phi^A)^* \phi^{A'} (\varphi_2^B)^* \varphi_2^{B'} \epsilon^{ABC} \epsilon^{A'B'C} \nn \\
&=& \frac{1}{4} (\phi^A)^* \phi^{A} (\varphi_2^B)^* \varphi_2^{B} - \frac{1}{4} (\phi^A)^* \phi^{B} \varphi_2^A (\varphi_2^{B})^*~, \nn \\
(F_2^C)^*F_2^C &=&\frac{1}{4} (\phi^A)^* \phi^{A} (\varphi_1^B)^* \varphi_1^{B} - \frac{1}{4} (\phi^A)^* \phi^{B} \varphi_1^A (\varphi_1^{B})^*~.
\eea
These terms in the potential give rise top the mass terms of $\phi$:
\bea
(F_1^C)^*F_1^C + (F_2^C)^*F_2^C = m^{AB} \phi^A (\phi^B)^*~,
\eea
where the mass matrix $m^{AB}$ can be determined by the second order derivative
\bea
m^{AB} &=& \frac{\partial^2}{\partial \phi^A \partial (\phi^B)^*}  \left[ (F_1^C)^*F_1^C + (F_2^C)^*F_2^C \right] \nn \\
&=& \frac{1}{4} \left[ \delta^{AB} (\varphi_1^C)^* \varphi_1^{C} - (\varphi_1^A)^* \varphi_1^{B} \right] + \left( 1 \rightarrow 2 \right)~.
\eea
The three eigenvalues of the mass matrix $m^{AB}$ are
\bea
m_1 =  \frac{1}{4} U, \quad m_2 = \frac{1}{8} \left( U + \sqrt{ U^2 - 16 (f^A)^* f^A} \right),\quad m_3 = \frac{1}{8} \left( U - \sqrt{ U^2 - 16 (f^A)^* f^A} \right)~. \nn
\eea
where $U$ is the sums of the quadratic casimirs
\bea
U = (\varphi_1^C)^* \varphi_1^{C} + (\varphi_2^C)^* \varphi_2^{C}~, 
\eea
and $f^A$ is the derivative of $W$ with respect to $\phi^A$ (which is zero because of $F$-terms):
\bea
f^A = \frac{\partial W}{\partial \phi^A} = \frac{i}{2} \epsilon^{ABC}  \varphi_1^B \varphi_2^C = 0~.
\eea
Thus, the mass eigenvalues are
\bea
m_1 =  \frac{1}{4} U, \quad m_2 = \frac{1}{4}  U, \quad m_3 = 0~. \nn
\eea
Indeed, two components of $\phi$ are massive (and each of them has mass $U/4$) and the other component is massless.

\subsection{A theory with genus $g$} \label{sec:genu1}
In this subsection, we give an argument that, for a theory with genus $g$, there is an unbroken $U(1)^g$ gauge symmetry at a generic point on the Kibble branch.  As a special case, in Appendix \ref{sec:genus1}, we show that for the tadpole theory ($g=1$), the unbroken gauge symmetry is $U(1)$.  

\begin{figure}[htbp]
\begin{center}
\includegraphics[height=2cm]{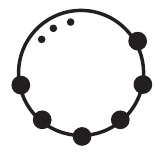}
\caption{For a loop, there are nodes on the loop.  Each node is attached to two lines.}
\label{formaloop}
\end{center}
\end{figure}
For a given loop, there are nodes and legs that go around it, as depicted in \fref{formaloop}.  Consider a node and two lines which are attached to it.  These two lines give rise to $V=6$ gauge fields in the vector multiplets ($3$ from each line).  The node itself gives rise to $H= 4$ hypermultiplets. 

After the Higgs mechanism, three gauge fields become massive and hence we are left with $V=3,~H=1$.  If the line is external, this $H=1$ hypermultiplet contributes to the dimension of the Kibble branch. Therefore, this effective process replaces a node with the two lines by a single line. Thus, one can keep eliminating nodes in such a way until the end result is a loop.  For such a loop, there is an unbroken $U(1)$ symmetry (from Appendix \ref{sec:genus1}).  One can proceed in this way for all loops, and concludes that for a theory with genus $g$, there is an unbroken $U(1)^g$ gauge symmetry at a generic point on the Kibble branch.

\end{document}